\pdfoutput=1
\PassOptionsToPackage{unicode}{hyperref}
\PassOptionsToPackage{hyphens}{url}
\documentclass[
  letterpaper,
  11pt]{article}
\usepackage{amsmath,amssymb}
\usepackage{lmodern}
\usepackage{iftex}
\usepackage[T1]{fontenc}
\usepackage[utf8]{inputenc}
\usepackage{textcomp}
\IfFileExists{upquote.sty}{\usepackage{upquote}}{}
\IfFileExists{microtype.sty}{%
  \usepackage[]{microtype}
  \UseMicrotypeSet[protrusion]{basicmath} %
}{}
\makeatletter
\@ifundefined{KOMAClassName}{%
  \IfFileExists{parskip.sty}{%
    \usepackage{parskip}
  }{%
    \setlength{\parindent}{0pt}
    \setlength{\parskip}{6pt plus 2pt minus 1pt}}
}{%
  \KOMAoptions{parskip=half}}
\makeatother
\usepackage{xcolor}
\IfFileExists{xurl.sty}{\usepackage{xurl}}{} %
\IfFileExists{bookmark.sty}{\usepackage{bookmark}}{\usepackage{hyperref}}
\hypersetup{
  pdfauthor={Adam Y. Shavit},
  hidelinks,
  pdfcreator={LaTeX via pandoc}}
\usepackage{longtable,booktabs,array}
\usepackage{calc} %
\usepackage{etoolbox}
\makeatletter
\patchcmd\longtable{\par}{\if@noskipsec\mbox{}\fi\par}{}{}
\makeatother
\IfFileExists{footnotehyper.sty}{\usepackage{footnotehyper}}{\usepackage{footnote}}
\makesavenoteenv{longtable}
\usepackage{graphicx}
\makeatletter
\def\maxwidth{\ifdim\Gin@nat@width>\linewidth\linewidth\else\Gin@nat@width\fi}
\def\maxheight{\ifdim\Gin@nat@height>\textheight\textheight\else\Gin@nat@height\fi}
\makeatother
\setkeys{Gin}{width=\maxwidth,height=\maxheight,keepaspectratio}
\makeatletter
\def\fps@figure{htbp}
\makeatother
\providecommand{\tightlist}{%
  \setlength{\itemsep}{0pt}\setlength{\parskip}{0pt}}
\usepackage[T1]{fontenc}
\usepackage[utf8]{inputenc}
\usepackage{lmodern}
\usepackage{textcomp}
\usepackage{amsmath,amssymb}
\usepackage{tipa}
\usepackage{booktabs,longtable,array}
\usepackage{graphicx}
\usepackage[margin=1in]{geometry}

\AtBeginDocument{\hypersetup{%
  pdftitle={Orthographic depth at two scales, with English also matched for lexicon size},%
  pdfauthor={Adam Y. Shavit}}}

\title{Orthographic depth at two scales, with English also matched for lexicon size}

\DeclareUnicodeCharacter{27E8}{\ensuremath{\langle}}
\DeclareUnicodeCharacter{27E9}{\ensuremath{\rangle}}
\DeclareUnicodeCharacter{026A}{\textipa{\textsci}}
\DeclareUnicodeCharacter{025B}{\textipa{E}}
\DeclareUnicodeCharacter{0124}{\^{H}}
\DeclareUnicodeCharacter{2082}{\ensuremath{_2}}
\DeclareUnicodeCharacter{03A3}{\ensuremath{\Sigma}}
\DeclareUnicodeCharacter{2212}{\ensuremath{-}}
\DeclareUnicodeCharacter{2192}{\ensuremath{\rightarrow}}
\DeclareUnicodeCharacter{00D7}{\ensuremath{\times}}
\DeclareUnicodeCharacter{0304}{\ensuremath{\bar{}}}
\DeclareUnicodeCharacter{0302}{\ensuremath{\hat{}}}
\DeclareUnicodeCharacter{2265}{\ensuremath{\geq}}
\DeclareUnicodeCharacter{02D0}{\textipa{:}}
\DeclareUnicodeCharacter{03A0}{\ensuremath{\Pi}}
\DeclareUnicodeCharacter{2016}{\ensuremath{\Vert}}
\DeclareUnicodeCharacter{207B}{\ensuremath{^{-}}}
\DeclareUnicodeCharacter{00B9}{\ensuremath{^{1}}}
\DeclareUnicodeCharacter{220E}{\ensuremath{\blacksquare}}
\DeclareUnicodeCharacter{2013}{\textendash}

\ifLuaTeX
  \usepackage{selnolig}  %
\fi

\author{Adam Y. Shavit}
\date{}

\begin{document}

\hypertarget{orthographic-depth-at-two-scales-with-english-also-matched-for-lexicon-size}{%
\section{Orthographic depth at two scales, with English also matched for
lexicon
size}\label{orthographic-depth-at-two-scales-with-english-also-matched-for-lexicon-size}}

\textbf{Adam Y. Shavit}

\emph{Hunter College and the Graduate Center, City University of New
York}\\
ORCID 0009-0008-1235-0995 · Correspondence: as1127@hunter.cuny.edu

\begin{center}\rule{0.5\linewidth}{0.5pt}\end{center}

\hypertarget{abstract}{%
\subsection{Abstract}\label{abstract}}

A reader can struggle to read in only one of their languages, and one
proposed reason is how predictably spelling maps to sound. We measure
that predictability as conditional entropy, the pronunciation
uncertainty left once the spelling is seen, at two grain sizes (unit
sizes): the \textbf{alignment unit}, an aligned spelling-sound chunk,
and the \textbf{syllable}, bounded in the pronunciation. We analyse word
lists in twelve alphabetic orthographies and report five in full, with
English pronunciations from a dictionary and the rest from the
synthesizer eSpeak-ng, length marks removed. Among the five, English has
the most uncertainty and Finnish the least at both grains. Across all
twelve, Portuguese and Danish exceed English per alignment unit, and
German and Italian order differently by grain and transcription. These
orders hold when English is transcribed by eSpeak-ng with u allowed to
spell /j/ plus a vowel. Uncertainty per unit has \textbf{no} shared
direction as units grow: it falls in Spanish and, less securely, German,
is level in Finnish, depends on transcription in Italian and on the word
list in English, and rises in Norwegian and Swedish. Summed uncertainty
per word falls in all twelve, but that score is handed the
pronunciation's boundaries and mixes context with in-sample
memorisation, so it is no evidence of what a reader computes. An exact
identity states when discarded spelling boundaries could reverse that
fall. A uniform 20,000-word English sample holds 35\% of the
dictionary's syllable spellings and has syllable entropy 0.098 bits
lower, so lexicons must be matched in size. In an exploratory check over
eleven languages, syllable-scale entropy tracked first-year reading
accuracy (rho = \ensuremath{-}0.74, falling to \ensuremath{-}0.25 if
English is given the lowest entropy instead), without separating the
syllable scale from the alignment unit. These estimates describe sampled
writing systems, not readers.

\hypertarget{introduction}{%
\subsection{1. Introduction}\label{introduction}}

A reader can have reading difficulty in one of their languages and not
in another. AS (Wydell \& Butterworth, 1999), a 16-year-old
English-Japanese bilingual, had reading and writing difficulties
confined to English, while his reading of Japanese kanji and kana was
equivalent to that of Japanese undergraduates or graduates (p.~273). To
account for the dissociation Wydell and Butterworth proposed a
``hypothesis of granularity and transparency'': an orthography whose
print-to-sound mapping is one-to-one, or whose smallest unit
representing sound is coarse (a whole character or word), should not
produce a high incidence of phonological dyslexia (p.~280). Read
probabilistically, its transparency dimension concerns how predictably
spelling maps to sound. That predictability is a property of the writing
system, and it can be estimated from a word list. Its granularity
dimension concerns the size of the unit over which the mapping is read.

Orthographic depth, how consistently a writing system's spelling
represents its pronunciation, is a central variable in cross-language
research on learning to read and on developmental dyslexia. Languages
that differ in it differ in first-year word-reading accuracy
(differences that Seymour et al., 2003, attribute to orthographic depth,
with syllabic complexity affecting decoding), in the weight phonological
awareness carries as a predictor of reading (Ziegler et al., 2010) and,
on one account, in children's reading strategies and the form dyslexia
takes (Ziegler \& Goswami, 2005). At least one study that needed depth
as a number found the available measures wanting. Landerl et al.~(2013),
comparing predictors of dyslexia across six orthographies in the
NEURODYS study, judged the available entropy figures insufficiently
valid to use as a continuous predictor and fell back on a three-way
complexity classification. Their reason was that Borgwaldt's entropy
measure is ``based only on word onsets'' and so ``vastly underestimates
some language differences'' (p.~688), missing irregularity elsewhere in
the word.

\textbf{Entropy measures of depth.} Entropy, here the uncertainty left
in a pronunciation once its spelling is known, described
spelling-to-sound mappings before it was used to compare many languages.
Treiman et al.~(1995) computed it for the onsets, nuclei, codas, bodies
and rimes of English CVC monosyllables, and Martensen et al.~(2000)
replicated that analysis for Dutch (both as cited in Borgwaldt et al.,
2004, p.~168; Lange and Content, 1999, p.~438, likewise name Treiman et
al.~as earlier users of the entropy statistic). Lange and Content (1999,
pp.~437--438) computed the entropy of each French grapheme's
pronunciation over 18,510 words of one to three syllables. Kessler and
Borgwaldt (in preparation, as reported in Borgwaldt, 2003, §6.6, p.~105)
computed grapheme entropy for 23,250 English words, scoring each word by
the sum of its graphemes' entropies. Borgwaldt, Hellwig and De Groot
(2004, 2005), applying a method of Smith and Silverberg's (as cited in
Borgwaldt et al., 2004, pp.~168--169), took entropy across five
languages and then seven as \emph{onset entropy}: uncertainty in mapping
a word's first letter to its first phoneme. That onset restriction is
the one Landerl et al.~objected to. Borleffs, Maassen, Lyytinen and
Zwarts (2017), reviewing regularity, consistency and entropy measures of
transparency, likewise note that word-initial entropy ``may still fail
to represent the full spectrum of potential mapping complexities in
different parts of the word'' (their Discussion).

\textbf{The unit is a choice, and the choice moves the estimate.}
Protopapas and Vlahou (2009) computed entropy for Greek in both
directions, over every position in the word and over word onsets, with
the reading direction at both the letter and the grapheme level (their
Table 4), exploiting Greek's complete sequential alignability. Their
feedforward consistency, the proportion of mappings that agree with the
dominant spelling-to-sound correspondence, is 95.1\% at the grapheme
grain and 80.3\% at the letter grain (token counts, reading direction).
The choice of unit alone moved one language's estimate in one direction
by 14.8 percentage points, which is direct published evidence that a
transparency estimate is not invariant to grain. They report it as a
methodological note: ``The grapheme unit size was selected as most
appropriate because the corresponding calculation, using single letters
instead of graphemes, resulted in a very substantially lower consistency
estimate (80.3\%) and a greater number of mappings (173) in the reading
direction'' (p.~996). They treat the unit as a decision with a right
answer. We treat grain as a measurement parameter to be varied
explicitly.

Others have varied the unit within the syllable. Siegelman, Kearns and
Rueckl (2020) built a fuller framework for English monosyllables.
Holding the unit fixed at the vowel grapheme and varying only the
context supplied, they report 0.69 bits with no neighbouring context,
0.37 bits when the onset is given and 0.25 bits when the coda is given.
Additional context can therefore cut residual uncertainty by nearly two
thirds without the unit changing at all. Lim, O'Brien and Onnis (2024)
compared consistency, in monosyllabic words, across several sub-syllabic
English units (grapheme, onset, vowel, coda, rime) that had previously
been studied in isolation. In monomorphemic Dutch and German words of
one or more syllables, Borgwaldt and De Groot (2002, pp.~54--56) scored
the consistency of overlapping clusters of onsets, nuclei and codas that
cross syllable boundaries. They set aside analysing each syllable
separately because an ambisyllabic consonant, which closes one syllable
and opens the next (the /m/ of German ⟨Hummer⟩), blurs the boundary
(p.~52). English consistency norms already syllabify multisyllabic
words, by the maximum onset principle, and score five segments within
each syllable (onset, nucleus, coda, oncleus and rime) position by
position. Their composite for a multisyllabic word is a segment's mean
consistency across its syllables, so no measure is defined on the whole
syllable, and they cover one language and measure consistency rather
than entropy (Chee, Chow, Yap \& Goh, 2020, pp. 2540–2541).

Schmalz, Rueckl and Siegelman (2026) provide the nearest comparison.
Across eight European orthographies they compute consistency at two
grain sizes (the vowel grapheme and the body/rime) and vowel entropy
with and without the neighbouring onset or coda as context. Over
letters, bigrams and graphemes aligned at every position, they also
compute the conditional entropy of a phonological unit given its
orthographic unit, which falls in all eight languages when the unit
grows from the letter to the bigram (their Table 5, pp.~10--11). They
argue that unit size and mapping systematicity are separable dimensions
of depth. They restrict their main cross-language comparisons to
monosyllabic words and state that extending to polysyllabic words would
exceed their scope. They also raise two problems any such comparison
inherits. Their corpora range from 1,155 to 8,027 monosyllabic words
across languages, a roughly sevenfold spread, large enough that observed
unit counts and the inclusion of rare word types are not directly
comparable. And apparent orthographic irregularity can instead come from
phonological context rules: in Russian, a language outside their sample,
consonant devoicing and palatalization would inflate an automated
estimate such as onset entropy although they are predictable from
phonology (p.~4), a confound automated entropy cannot separate without
inspection. Siegelman et al.~(2020, p.~1309), in their Discussion, name
the cross-writing-system extension as their own scope limit, ``here we
only examined a single writing system -- English'', and hand it on:
``future cross-linguistic studies are left with the task of comparing
the information distributions of orthography-to-phonology of different
writing systems'', citing Borgwaldt et al.~and Ziegler et al.~as the
restricted-input work their metrics could generalise.

\textbf{Whole-word and computational measures.} A second line of work
scores depth over the whole word. Van den Bosch, Content, Daelemans and
De Gelder (1994) compared Dutch, English and French by how well
letter-to-phoneme models trained on each generalise to unseen words,
storing for each letter the minimal context that makes its pronunciation
unambiguous and ordering the context letters by their information gain,
an entropy reduction (pp.~5 and 8 of the authors' preprint). Marjou
(2021, p.~1) trained a neural network on whole words to predict their
readings and spellings in 17 orthographies. Schmalz, Marinus, Coltheart
and Castles (2015, abstract) propose that orthographic depth is ``a
conglomerate of two separate constructs: the complexity of
print-to-speech correspondences and the unpredictability of the
derivation of the pronunciations of words on the basis of their
orthography''. Torres and Futrell (2025) measure orthographic
transparency as compressibility across 22 languages, with one input
representation (characters, with sub-character components split out for
Chinese, Japanese and Korean) that they do not vary across grain sizes.
They do control sample size: they keep the 5,032 most frequent forms per
language, ranked on a separate news corpus, because ``highly infrequent
forms might be nonrepresentative in terms of orthography and
phonotactics'' (their §4.1). These measures count rules, score
consistency or score a trained model's accuracy. Van den Bosch et al.'s
information gain is an entropy measure, but it varies the context around
a fixed one-letter unit and serves to order a decision tree. None of
these whole-word works reports a conditional entropy at a unit size the
analysis varies. In computational linguistics the immediate context is
the SIGMORPHON 2022 shared task (McCarthy et al., 2023), which
benchmarks grapheme-to-phoneme conversion across ten related,
script-sharing language pairs. The task used the same languages in every
setting so that its results could be compared, chose them ``to avoid
particularly pathological languages noted in previous years (English,
Croatian)'' and phenomena such as Danish stød, and omitted English data
over quality concerns (McCarthy et al., 2023, p.~231).

\textbf{This paper.} We measure how predictably spelling maps to sound
in twelve alphabetic orthographies and report five of them in full:
English, German, Italian, Spanish and Finnish. The measure is
conditional entropy, $H(P\mid G)$, the entropy of a unit's pronunciation
$P$ given its spelling $G$. We compute it over every aligned unit in
every word of full polysyllabic lexicons, at two grain sizes, the
\textbf{alignment unit} (a spelling chunk and its pronunciation) and the
\textbf{syllable}, and we match English to the other lexicons for size.
A measure's \textbf{grain size} here is the size of the aligned
spelling--pronunciation unit over which the conditional distribution is
computed. It is not the amount of surrounding context supplied to a unit
of fixed size, as in Siegelman et al., and not the reader-side grain of
psycholinguistic grain size theory (Ziegler \& Goswami, 2005). Each
estimate describes a writing system as sampled by a particular word list
and transcription. The paper reports three results. English is highest
and Finnish lowest of the five at both grains (§3.1). Per-unit entropy
has no common direction as units grow from the alignment unit to the
syllable (§3.2). Lexicon size changes the syllable-grain estimate: a
uniform 20,000-word sample of the English dictionary holds 35\% of its
syllable spellings and has syllable entropy 0.098 bits lower, on average
over 1,000 draws (§3.3).

The work adds three things to the literature above.

\begin{enumerate}
\def\labelenumi{\arabic{enumi}.}
\tightlist
\item
  \textbf{All-position conditional entropy over full polysyllabic
  lexicons.} Each aligned unit in each word contributes to the measure,
  so it covers the irregularity outside the word onset, which Landerl et
  al.~found onset entropy to miss and which Borleffs et al.~note it may
  fail to represent. Covering that irregularity is not the same as
  predicting reading better (§3.5). ``All-position'' means that every
  unit contributes, not that the quantity is a whole-word conditional
  entropy, which it is not (§2). Protopapas and Vlahou computed this
  all-position entropy for Greek letters and graphemes, Lange and
  Content for French graphemes over words of up to three syllables,
  Kessler and Borgwaldt for English graphemes and Schmalz et al.~for the
  monosyllables of eight languages. We compute it over full polysyllabic
  lexicons in five languages with one aligner. A measure defined over
  the whole word is the one every grain-size result below is computed
  on, because a syllable grain has no onset-only counterpart.
\item
  \textbf{The syllable grain over polysyllabic vocabulary.} Prior work
  that varies the unit uses monosyllables (Treiman et al.~and Martensen
  et al., as cited above; Lim et al.; Schmalz et al.). Protopapas and
  Vlahou vary the letter against the grapheme over every position in
  Greek words, and Borgwaldt and De Groot score clusters that cross
  syllable boundaries. Within the work we located, none of these takes
  the whole syllable as the scored unit. We syllabify full polysyllabic
  lexicons in five languages and take the syllable itself as the unit,
  which is the extension Schmalz et al.~explicitly decline. Our
  syllabifier handles the ambisyllabic consonant that Borgwaldt and De
  Groot set aside by one convention, giving it and all its letters to
  the following syllable (§2), so ⟨Hummer⟩ divides as ⟨Hu⟩ + ⟨mmer⟩, and
  §3.4 measures how much that convention matters.
\item
  \textbf{The lexicon-size control, applied rather than flagged, and one
  measured transcription artefact.} Schmalz et al.~identify corpus-size
  spread as a distortion. We match on it, and for English on lexicon
  composition too, and report English both matched and unmatched (§3.3;
  Table~\ref{tab:depth} is unmatched), because $N$ is far from
  saturating at the syllable grain. Our control differs from that of
  Torres and Futrell. Truncating by frequency rank removes the rare
  tail, and a uniform random sample of word types only thins it. §3.3
  and §4.4 concern that rare tail: 38\% to 59\% of observed
  syllable-spelling types occur exactly once in the full lexicon (in
  these five lexicons, equivalently, in a single word type), so the two
  controls measure different things, and they give different answers
  here. Four of our five lexicons are themselves frequency truncations,
  so §3.3 matches English both ways and prints both. Schmalz et al.~also
  name the phonotactic confound (pronunciation predictable from
  phonological context rather than from spelling) in general terms. We
  find one instance in our own data, produced by the transcription tool
  rather than by Finnish phonology, and measure it. eSpeak-ng
  transcribes the Finnish grapheme ⟨i⟩ as /ɪ/ in 72\% of its occurrences
  in the word list and /i/ in 28\%, and Finnish has one /i/ phoneme
  (Suomi, Toivanen \& Ylitalo, 2008, p.~20), so the transcription
  creates two output categories where the analysis intends one. We merge
  the two throughout and report what the merge changes (§3.4). Italian
  shows the same split, but its status there is not established, so we
  report Italian both ways. We release the candidate list, the merge and
  the unmerged transcription so the correction is auditable.
\end{enumerate}

\hypertarget{method}{%
\subsection{2. Method}\label{method}}

\textbf{Lexicons.} English from CMUdict 0.7a (Carnegie Mellon
University, 2008; 113,724 of 114,954 words aligned); German, Italian,
Spanish and Finnish phonemized with eSpeak-ng (eSpeak NG contributors,
2024) over 20,000-word frequency lists each. The lexicons therefore
differ in composition as well as size: English's is a whole pronouncing
dictionary, two thirds of whose aligned words are absent from the 50,000
most frequent words of the OpenSubtitles English list (Dave, 2018;
Lison, Tiedemann \& Kouylekov, 2018), the source of the other four
lists, while each of the other four is the most frequent 20,000 words of
its list. §3.3 matches English on both. Matching how a list was built
does not match its words' morphology, length or syllable structure,
which still differ across languages (Table~\ref{tab:depth} gives
syllables per word). The aligner drops the words it cannot align: 1,230
in English, almost all with more phonemes than letters (MUSIC, HUMAN,
the ⟨u⟩ → /ju/ class), and 75 to 129 in each of the other four. Before
alignment we remove CMUdict's stress digits and eSpeak-ng's stress
marks, combining diacritics and length mark (the IPA symbol ː), so a
long and a short sound share one symbol, which merges length contrasts
(vowels and consonants alike: eSpeak-ng marks the long /t/ of Italian
⟨notte⟩ with it), and every letter and phoneme belongs to exactly one
unit (a silent letter forms a unit with no phoneme or joins a
many-to-one unit); §3.1 reports what keeping the length mark changes.
The eSpeak-ng version that transcribed the eleven non-English lists was
not recorded; for German it was a release older than 1.52.0, which wrote
one phoneme, the vowel of ⟨ur⟩ (wurde, durch), as a placeholder of two
question marks in 201 words; we replace it with the symbol eSpeak-ng
1.52.0 writes for that phoneme, which gives those syllables their vowel
and leaves every alignment's segmentation unchanged. A scan of every
phoneme token in the twelve languages' alignments finds no other
placeholder apart from one token in one German word. \textbf{The
pronunciation sources are therefore not matched}, and we state the
asymmetry rather than claiming it away: English pronunciations are
lexicographic, the other four are rule-generated, and §3.4 shows that
the transcription layer can move an entropy estimate by 69\% of its
unmerged value (Finnish). A source-matched check, defined after the
alignment paragraphs below and reported in §3.4, measures that confound
for English end to end.

\textbf{Sampling distribution.} Each lexicon is a list of word types,
and every type contributes once; the averages below are over aligned
unit tokens (each occurrence of a unit in the list, every word counted
once) pooled across types, not over corpus frequency. Estimates weighted
by how often words occur in running text are not reported, and
``corpus'' below means this word list.

\textbf{Alignment.} An aligner trained by EM (expectation maximization,
an iterative method for fitting probabilities when alignments are
unknown) over units that pair several letters with one phoneme, with one
letter spelling several phonemes allowed only for ⟨x⟩ (and ⟨z⟩ in
German)~--- forward--backward over an alignment lattice, then Viterbi
decoding~--- implemented using only the Python standard library; every
table below counts each word's single Viterbi alignment. An alignment is
a sequence of paired spelling and pronunciation chunks; each candidate
pair is an edge in the lattice. Two constraints apply: an edge may not
contain multiple letters \emph{and} multiple phonemes at once, and only
the listed letters may spell more than one phoneme. Without the first
the likelihood favours unnaturally large chunks, because a segmentation
with fewer multiplicative sub-one factors can receive a higher path
probability, and EM collapses toward whole-word chunks. The restriction
yields units that resemble familiar grapheme--phoneme correspondences
while still allowing the one-to-many case ⟨x⟩ → /ks/. \textbf{It is a
modelling choice, not a definition of a grapheme}, and we do not present
it as one. EM and EM-like estimation of letter-to-phoneme alignments is
established work (Black, Lenzo \& Pagel, 1998; Damper et al., 2005), as
are non-one-to-one correspondence units (Bisani \& Ney, 2003), with
Galescu and Allen (2001) and Jiampojamarn, Kondrak and Sherif (2007)
combining both: multi-letter, multi-phoneme units inferred by EM, the
second by forward--backward training, as here. Ours is a self-contained
reimplementation of that family, chosen so the pipeline has no build
dependencies.

\textbf{Settings.} English: CMUdict 0.7a as distributed with NLTK,
keeping lower-case alphabetic entries of at most 14 letters (114,954 of
115,533 alphabetic entries) with the first listed pronunciation. Other
languages: the most frequent 20,000 entries of each OpenSubtitles 2018
list that are written in that language's letters and are 2 to 14 letters
long, phonemized by eSpeak-ng with \texttt{-\/-ipa\ -\/-sep=\_}, so a
phoneme is whatever token eSpeak-ng emits between separators: an
affricate such as German /ts/ or a diphthong is one token, and so one
syllable nucleus at most. Voices: \texttt{de}, \texttt{it}, \texttt{es}
and \texttt{fi} for the four; \texttt{hu}, \texttt{nl}, \texttt{pt}
(European Portuguese, not \texttt{pt-br}), \texttt{fr-fr}, \texttt{da},
\texttt{sv} and \texttt{nb} for the other seven. EM runs 8 iterations
with no convergence test, from initial weights spread over every unit
pair the lattice allows with one-letter-to-one-phoneme pairs weighted 20
to 1; a unit holds at most 4 letters and at most 2 phonemes. English,
the only lexicon above 25,000 words, is trained on a random 25,000-word
subsample (seed 31) and every word is then decoded; the uniform
20,000-word English sample of §3.3 is also drawn with seed 31, from the
aligned words, and keeps their alignments. The sonority scale is stop
\textless{} affricate \textless{} fricative \textless{} nasal
\textless{} liquid \textless{} glide \textless{} vowel. Every symbol
eSpeak-ng emits is assigned to one of these classes; two needed care.
eSpeak-ng writes one Danish vowel with the Greek letter epsilon, which
the syllabifier counts as a vowel, and the labiodental approximant it
writes for Danish ⟨v⟩ and Dutch ⟨w⟩ ranks as a glide. Finnish ⟨i⟩ is
written with two symbols, /ɪ/ and /i/, which we merge into one (§3.4);
the merge relabels symbols and leaves every alignment's segmentation
unchanged.

\textbf{Alignment check.} A uniform random sample of 300 English and 200
German aligned words (\texttt{run\_alignment\_audit.py}, seed 20261020)
was judged word by word against a stated criterion by a separate Claude
instance that had not seen the aligner's code or this paper's results:
an alignment is correct when every unit pairs letters with the phonemes
those letters spell in that word, and, where two placements are both
standard (a silent ⟨e⟩ joined to the vowel or left on its own; a doubled
consonant's second letter silent or merged), either is accepted.
\textbf{19 of 300 English words (6.3\%, Wilson 95\% interval 4.1\% to
9.7\%) and 3 of 200 German words (1.5\%, 0.5\% to 4.3\%) are
misaligned.} Of the English 19, 14 misplace a boundary between
neighbouring units without moving a phoneme past its letter (the schwa
of ⟨-le⟩ on the ⟨l⟩, a vowel letter joined to a consonant); 5 move
phonemes onto letters that do not spell them. The check estimates how
often an alignment departs from that criterion; it is one model reader's
application of standard grapheme--phoneme analysis (a second Claude
session re-judged all 22 errors and 13 of the correct words and agreed
on each), not an annotated corpus or a second aligner, and it does not
by itself say how far the misalignments move an entropy.

\textbf{A forced misalignment, measured.} Only ⟨x⟩ (and German ⟨z⟩) may
spell two phonemes, but English ⟨u⟩ often spells two, /j/ and /u/
(CUBE). If such a word has as many letters as phonemes, the only
alignment the constraints allow gives ⟨u⟩ the /j/ and moves every later
phoneme one letter on (ABUSE pairs ⟨s⟩ with /u/ and ⟨e⟩ with /s/); in
some other words a consonant letter takes the /j/ (ACCU). This touches
1,503 English words (1.3\% of the aligned dictionary's words, which hold
1.8\% of its unit tokens) and no other language. Re-running the aligner
with one added permission, that ⟨u⟩ may spell /j/ and a following vowel
as one unit, lowers English $H_{\mathrm{align}}$ by 0.042 to 0.046 bits
(0.936 to 0.890 over the same words, to 0.894 over all the words it
aligns) and $H_{\mathrm{syll}}$ by 0.002 to 0.008 bits. Every ranking
reported below holds under that rule, but one sign does not: in the
frequency-matched English list the per-unit change from alignment unit
to syllable becomes a 0.035-bit rise instead of a 0.015-bit fall (§3.2,
§3.3). The tables keep the alignment specified above;
\texttt{run\_ju\_check.py} reports both alignments, with a control that
retraining without the rule reproduces the released English alignment
exactly.

\textbf{A source-matched English check.} We also transcribed the same
114,954 English words with eSpeak-ng, so that the intended change is the
transcription process alone. The comparison is still an end-to-end
source sensitivity, not an isolated source effect: the aligner is
retrained on each transcription, and the set of words it aligns differs.
The voice is \texttt{en-us}, because CMUdict transcribes General
American. We aligned the transcriptions with the settings above (113,454
aligned). The syllabifier's vowel set gains the three vowel symbols only
eSpeak-ng's English emits, a reduced vowel and two r-coloured vowels.
eSpeak-ng also writes the /ju/ of ⟨u⟩ as two phonemes, so the forced
misalignment measured above is in its alignment too, in 2,002 words. We
therefore use as the source-matched check the eSpeak-ng run with the
/ju/ permission defined above (114,353 words aligned under it). We
compare it with the dictionary under the same permission, and §3.4
reports the comparison. The tables keep the dictionary.
\texttt{run\_espeak\_english.py} transcribes, aligns and reruns every
deposited script on both alignments, with a control that the rerun
reproduces the released outputs exactly on the dictionary alignment.
\texttt{run\_ju\_check\_espeak.py} applies the /ju/ permission to the
eSpeak-ng alignment and reruns the scripts behind these claims, with a
control that retraining without the permission reproduces the eSpeak-ng
alignment exactly. Transcribing needs eSpeak-ng, so the transcriptions
and all three eSpeak-ng English alignments (the one without the
permission and the two that \texttt{run\_ju\_check\_espeak.py} re-aligns
under it) ship in the deposit.

\begin{figure}[!htbp]
\centering
\refstepcounter{figure}\label{fig:decomp}
\includegraphics[width=0.92\textwidth,height=0.7\textheight]{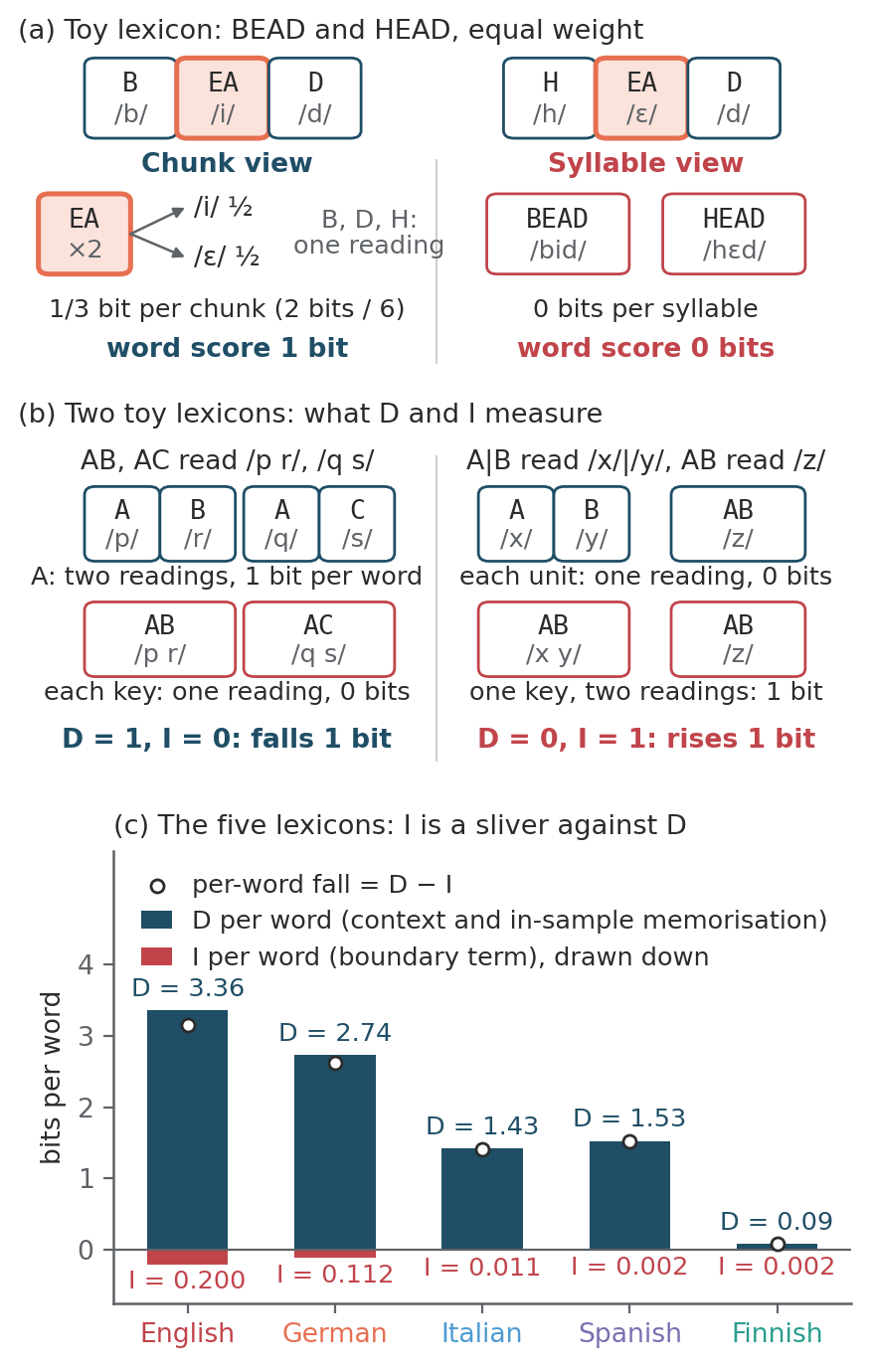}
\par\medskip
\begin{minipage}{\linewidth}
\textbf{Figure 3}~--- The per-word decomposition of §3.2 and its
spelling-boundary term. (a) Two made-up words, BEAD read /bid/ and HEAD
read /hɛd/. Chunk by chunk, EA has two readings, so alignment-unit
entropy is 1/3 bit per chunk and the unit score 1 bit per word. As whole
syllables, each spelling has one reading, so both syllable measures are
0. The toy shows what each measure counts, not how a reader segments
print. (b) The two toy lexicons of §3.2 and Appendix A, exact. Left: in
AB and AC the unit A has two readings, so the unit score is 1 bit per
word, while each syllable has one reading; the context the syllable
takes in is $D=1$ bit, $I$ is 0, and the score falls. Right:
A\textbar{}B and the single unit AB are each read one way, but the
syllable spelling AB (the key in panel b), which forgets where the unit
boundary was, has two readings; $D$ is 0, $I$ is 1 bit, and the score
rises. Only $I$ can raise the score. (c) $D$ and $I$ per word in the
five lexicons, from \texttt{run\_grain\_decomposition.py}. The dot is
the per-word fall, $D$ minus $I$, because $H(Y\mid P,X)$ is zero in all
five. $I$'s lexicon mean is largest in English, 0.20 bits per word. $D$
is in-sample and inherits the plug-in's sparse-observation sensitivity
§4.4 describes.
\end{minipage}
\end{figure}

\textbf{Syllabification.} We first identify vowel nuclei on the
pronunciation side. Intervocalic consonants are then assigned to the
following syllable whenever that produces an onset the
sonority-sequencing heuristic allows (``maximal onset''), with a
separate rule for /s/-clusters and a ban on splitting an aligned unit.
Boundaries fall between units, so a syllable is a concatenation of
alignment units and syllable-scale entropy is computed exactly as
alignment-unit entropy is. The spelling of such a syllable is what we
call a \emph{syllable spelling}: the letters of a syllable whose
boundaries are found on the pronunciation side, not a spelling-defined
unit such as the orthographic syllables of reading research. A reader
who sees only the printed word is not guaranteed to recover a syllable
spelling, because its boundaries are fixed with the pronunciation in
hand. Figure~\ref{fig:decomp} labels it the key for short.

\textbf{Entropy.} At grain $g$, spelling unit $x$ has probability
$p(x)$; its aligned pronunciation $y$ has conditional probability
$p(y\mid x)$. The average pronunciation uncertainty is

\[
H_g=-\sum_x p(x)\sum_y p(y\mid x)\log_2 p(y\mid x),
\]

over aligned unit tokens, where $x$ ranges over spelling units and $y$
over the pronunciation units aligned to them. In words: $H_g$ is the
expected number of bits of pronunciation uncertainty that remain once
the spelling unit is known, averaging over spelling units in proportion
to how often they occur in the analysed lexicon. It is a
\textbf{per-unit} quantity at the chosen grain, and it is not the
conditional entropy of a whole word's pronunciation given its spelling.
Computed, as throughout, from observed proportions, $H_g$ is the
\emph{plug-in} estimate, written $\hat{H}_g$ where its role as an
estimate matters (§3.2).

\textbf{\boldmath $N$} is the number of distinct spelling units observed
at least once in the analysed lexicon~--- \textbf{observed support}, not
a census of the units a reader has stored. We reserve ``inventory'' for
an explicitly qualified phrase.

\textbf{The entropies are in-sample corpus descriptions.} The alignment
probabilities are estimated from the same lexicon on which entropy is
then computed, so every figure below describes the analysed word list
rather than predicting unseen words. The average is also weighted by how
many times each spelling unit occurs across the list's words, so $H_g$
is a property of a writing system \emph{as sampled by this word list}
and not a corpus-free constant of the orthography. Held-out conditional
cross-entropy is the quantity a predictive claim would need, and §4.4
says why the need is greatest at the syllable scale.

\textbf{Reading direction only.} The released alignments also permit
$H(G\mid P)$, which we leave to future analysis.

\hypertarget{results}{%
\subsection{3. Results}\label{results}}

\hypertarget{depth-at-two-grains}{%
\subsubsection{3.1 Depth at two grains}\label{depth-at-two-grains}}

\begin{table}[htbp]
\centering
\caption{Conditional entropy of pronunciation given spelling and the
number of observed spelling types, at the alignment-unit and syllable
grains, with syllables per word, for the five reported languages
(English from the dictionary, unmatched for lexicon size and aligned
without the /ju/ permission of §2; Finnish with eSpeak-ng's /ɪ/ and /i/
merged, Italian as eSpeak-ng writes it, with merged values in
§3.4).}\label{tab:depth}
\begin{tabular}{@{} >{\raggedright\arraybackslash}p{(\columnwidth - 10\tabcolsep) * \real{0.1666}} >{\raggedright\arraybackslash}p{(\columnwidth - 10\tabcolsep) * \real{0.1666}} >{\raggedright\arraybackslash}p{(\columnwidth - 10\tabcolsep) * \real{0.1666}} >{\raggedright\arraybackslash}p{(\columnwidth - 10\tabcolsep) * \real{0.1666}} >{\raggedright\arraybackslash}p{(\columnwidth - 10\tabcolsep) * \real{0.1666}} >{\raggedright\arraybackslash}p{(\columnwidth - 10\tabcolsep) * \real{0.1666}}@{}}
\toprule
language & $H_{\mathrm{align}}$ (bits/alignment
unit) & $N_{\mathrm{align}}$ (types) & $H_{\mathrm{syll}}$
(bits/syllable) & $N_{\mathrm{syll}}$ (types) & syllables/word \\
\midrule
English & 0.936 & 532 & 1.103 & 27,162 & 2.43 \\
German & 0.604 & 174 & 0.510 & 6,252 & 2.46 \\
Italian & 0.432 & 182 & 0.510 & 2,690 & 3.13 \\
Spanish & 0.323 & 97 & 0.249 & 2,965 & 2.96 \\
Finnish & 0.029 & 68 & 0.039 & 2,889 & 2.95 \\
\bottomrule
\end{tabular}
\end{table}

Every value in Table~\ref{tab:depth} is generated directly from the
released analysis files. The full-lexicon rankings and the calibration
in this section also hold with English transcribed by eSpeak-ng, as the
other eleven languages are, under the /ju/ permission of §2; §3.4 gives
those values.

\begin{figure}[!htbp]
\centering
\refstepcounter{figure}\label{fig:1}
\includegraphics[width=0.92\textwidth,height=0.7\textheight]{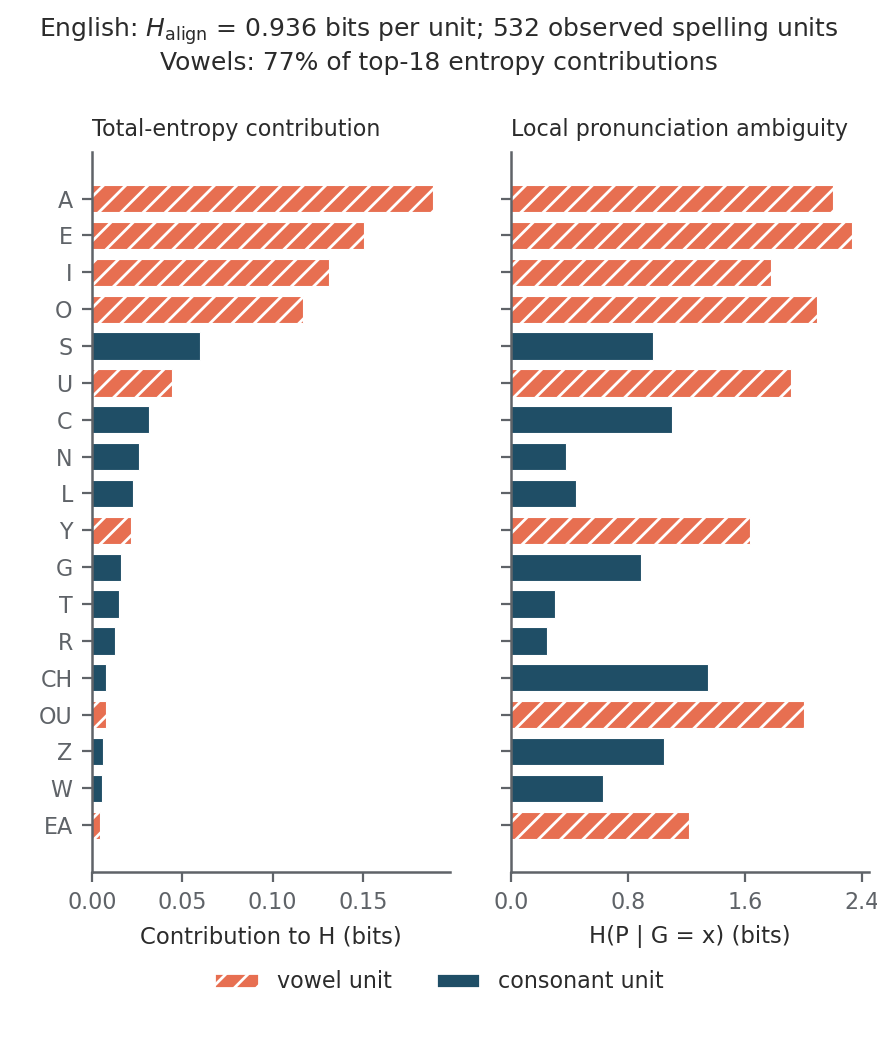}
\par\medskip
\begin{minipage}{\linewidth}
\textbf{Figure 1}~--- English's $H_{\mathrm{align}}=0.936$ bits, with
the 18 largest contributors shown among 532 observed spelling units.
Left: each unit's contribution to $H$, its share of the list's unit
tokens times its local entropy. Right: each unit's own local entropy
$H(P\mid G=x)$ (``local pronunciation ambiguity''), read in isolation.
Vowel units (spelled only with A, E, I, O, U or Y) carry 77\% of the
top-18 units' contribution; in the left panel, most consonant units
contribute little to $H$. Ranking is by contribution rather than by
local entropy, so rare but ambiguous units do not appear expensive.
Computed from the same EM-aligned lexicon as the table, adding no
parameters beyond the aligner settings of §2.
\end{minipage}
\end{figure}

\textbf{Among these five languages, the extremes are stable; the middle
is not.} English is highest and Finnish lowest at both grains
(Figure~\ref{fig:1} shows which English units carry its alignment
entropy). The full ordering is not identical: at the alignment-unit
grain German (0.604) sits above Italian (0.432), and at the syllable
grain the two are equal to four decimals (0.5104 bits each; Italian is
higher in the fifth decimal place). \textbf{A gap that small sits far
inside the 0.0249 bits the syllabification heuristic alone moves Italian
(§3.4), so the syllable-scale measure does not order German and
Italian}, while the alignment-unit measure puts German 0.17 bits above
Italian; the two grains do not agree, and we report no order that holds
at both grains. With Italian's /ɪ/ and /i/ merged as Finnish's are
(§3.4), German is higher at both grains (0.604 against 0.350, and 0.510
against 0.368). Restoring eSpeak-ng's length mark (§2) to the same
alignments puts German above Italian at both grains (0.558 against 0.530
at the syllable grain), with English still highest and Finnish lowest.
The pipeline's uppercasing also writes German ß as SS. Keeping the 275 ß
words apart lowers German's syllable entropy to 0.5094 bits, 0.0010
below Italian and still far inside Italian's 0.0249-bit syllabifier
spread, and leaves its alignment-unit value unchanged
(\texttt{run\_eszett\_check.py}). The same collapse-in-the-middle
pattern appears in the prior work: Borgwaldt et al.~(2004, p.~178)
report that ``at a grain size of three letters, English remains more
ambiguous than the other orthographies in the spelling-to-sound
direction, whereas French, Dutch, and German approximate one another''.
Their larger grain is the first two or three letters against the first
phoneme, so this is word-initial contextual entropy rather than entropy
at every position~--- the same collapse in the middle, in a narrower
measurement.

\textbf{Against published rankings, with the disagreements named.}
Caravolas (2022, p.~330) summarises a convergence across published
cross-linguistic comparisons~--- she calls the estimates between the two
extremes ``indicative'', and they characterise other studies rather than
report her own measurement. Scored against that summary, our
alignment-unit measure $H_{\mathrm{align}}$ orders 19 of 21 cross-band
language pairs in the same direction, with Spearman rho = −0.80 between
band index (1 = least consistent, 4 = most, so a negative rho is
agreement) and $H_{\mathrm{align}}$ over the 8 of our twelve languages
she places (English, Portuguese, Danish, German, Dutch, Italian,
Spanish, Finnish). French is scored apart because she places it in no
band, describing its feedback consistency as the lowest and its
feedforward consistency as ``moderately high''; our measure, in the same
feedforward (reading) direction, gives it the seventh-highest entropy of
twelve, below every language of her low-to-moderate band and above
Spanish and Italian. Shuffling the band labels across the same 8
languages 20,000 times (seed 20260807) puts the rho at one-sided p =
0.010 and the 19-of-21 pair count at one-sided \textbf{p = 0.014}; we
report each statistic with its own p-value. \textbf{Read these as
calibration, not as a hypothesis test} (the p-values say how often
shuffled bands do as well): n = 8, the bands are indicative, and Danish,
Dutch and German share a family and much of their orthographic history.
\textbf{Both violations are the same disagreement}: our measure puts
Portuguese (0.985) and Danish (0.980) above English (0.936), the reverse
of the summary's order. For Danish, published orders vary with the
measure: Schmalz et al.~(2026) place Danish above English on vowel
entropy but not on onset entropy (their Table 4), so agreement with us
depends on which measure is compared. For Portuguese, a bounded web
search of published depth, consistency and entropy rankings (October
2026; not a systematic review) found no published comparison that
agrees, and we report an open discrepancy rather than a result. Danish's
place depends on one transcription choice. eSpeak-ng marks Danish stød,
a glottal feature, on 18.6\% of Danish phoneme tokens, and deleting the
marker lowers Danish $H_{\mathrm{align}}$ from 0.980 to 0.829, below
English. Portuguese is then the only violation (20 of 21 pairs, rho =
−0.87, both one-sided p \textless{} 0.01). Every table keeps the marker,
because whether Danish spelling signals stød is a question this paper
does not settle.

\textbf{These five languages are a subset, and no principled rule picks
them out.} The pipeline computes twelve (these five plus Danish, Dutch,
French, Hungarian, Norwegian, Portuguese and Swedish) and the released
tables carry all twelve; the five reported here are the set this
analysis was built on, not a selection justified by a criterion. The
twelve-language results show why this selection matters. \textbf{In the
twelve-language table English is third at the alignment-unit grain},
behind Portuguese (0.985) and Danish (0.980) (English is second if
Danish's stød marker is deleted). At the syllable grain it is highest of
all twelve (1.103 bits, against Danish's 0.984). Matched on size alone
(1.005, Table~\ref{tab:size}) it is 0.021 bits above Danish, more than
either language's own spread across the five syllabifier settings of
§3.4 (0.004 and 0.011 bits over the full lexicons), and the central 95\%
of 1,000 further uniform draws (0.994 to 1.016 bits, §3.3) all lie above
Danish's 0.984. Matched on composition as well (0.879), it is 0.105 bits
below Danish, so we do not order the two. Finnish is lowest of the
twelve per alignment unit (0.0292 against Hungarian's 0.0465, an order
that holds only with Finnish's two i symbols merged, §3.4) and Hungarian
per syllable. We have measured no spread for alignment-unit differences,
because no syllabifier setting acts on them and we fitted the aligner
once (§4.4); the smallest gap we read as an order is Finnish's 0.017
bits below Hungarian, and we do not order the 0.0004-bit gap of §3.4.
Any criterion we could write down that admits these five~--- holding a
published outcome measure, a published depth band and a syllabification
check~--- admits Danish, Dutch and Portuguese as well, which is to say
it admits the languages that break the framing (Portuguese, and Danish
while its stød marker is kept). \textbf{Every claim about the extremes
and the full ordering in this section is therefore scoped to the five,
apart from the eight-language calibration above and the twelve-language
positions this paragraph states; the twelve-language table is released,
and the extremes should be read as a property of this five-language set
rather than of alphabetic orthographies in general.} The five are the
set the analysis was developed on, and the twelve-language table is the
full result. Agreement of either with published depth bands calibrates
the measure. It does not test the transparency hypothesis of §1, because
a language-level opacity label, language family or a simple
vowel-irregularity count could produce the same agreement.

\hypertarget{entropy-per-unit-has-no-common-direction-across-grain-sizes}{%
\subsubsection{3.2 Entropy per unit has no common direction across grain
sizes}\label{entropy-per-unit-has-no-common-direction-across-grain-sizes}}

\begin{figure}[!htbp]
\centering
\refstepcounter{figure}\label{fig:2}
\includegraphics[width=0.92\textwidth,height=0.7\textheight]{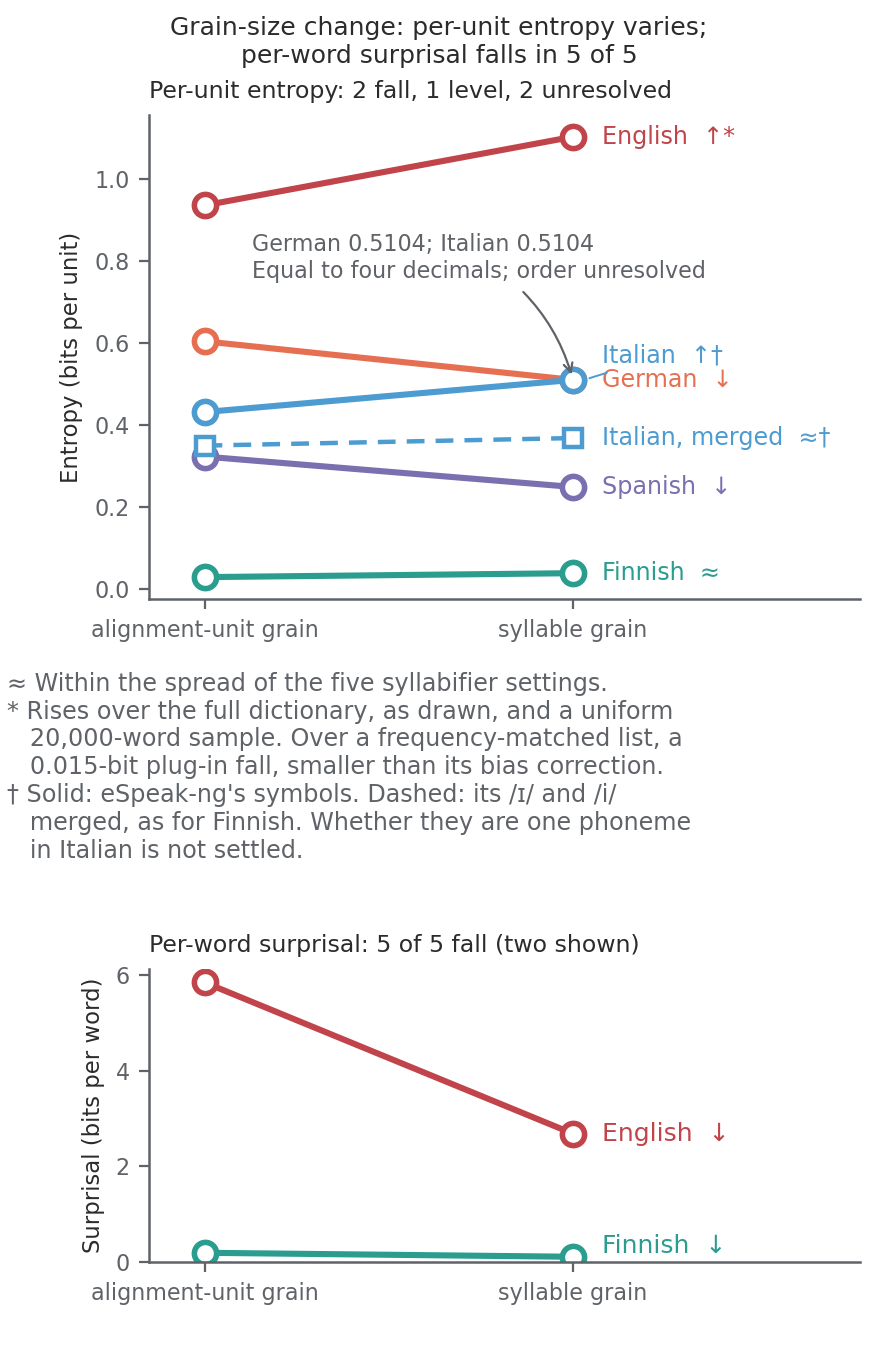}
\par\medskip
\begin{minipage}{\linewidth}
\textbf{Figure 2}~--- Top: all five languages' $H_{\mathrm{align}}$
beside $H_{\mathrm{syll}}$, one line per language, from
Table~\ref{tab:depth}, with Italian drawn twice: solid with eSpeak-ng's
symbols (the table's row) and dashed with its /ɪ/ and /i/ merged as
Finnish's are (§3.4). An arrow marks a direction only when its sign
holds under all five syllabifier settings of §3.4 and its size exceeds
that language's spread across them; $\approx$ marks a change within that
spread (Finnish, and Italian merged). The vertical axis is bits per unit
of each grain, per alignment unit at the left point and per syllable at
the right. All markers are drawn unfilled and at their unrounded
computed values; German and Italian are equal to four decimals, so their
markers overlap and the panel labels the tie rather than separating the
two points. English's rise holds over the full dictionary and a uniform
20,000-word sample but not over a frequency-matched list (where the
plug-in per-unit fall is smaller than its §4.4 bias correction, which
establishes neither a fall nor a rise), so the panel title counts
English, with Italian, as unresolved and English's label carries an
asterisk (§3.3). Bottom: additive conditional surprisal per word at each
grain, for the two languages whose per-word figures appear in prose
above; all five per-word falls are in Table~\ref{tab:decomp-terms} and
Figure~\ref{fig:decomp}.
\end{minipage}
\end{figure}

$H_{\mathrm{syll}}$ falls below $H_{\mathrm{align}}$ in German and
Spanish under all five syllabifier settings of §3.4, by more than each
language's spread across them (Figure~\ref{fig:2}); German's fall is the
less secure of the two. It keeps its sign under all three bias
corrections (Miller–Madow, a delete-one-word jackknife and extrapolation
to an unlimited lexicon), but when a simulated rise is planted in
German's own syllables, none of them reads it as a rise beyond German's
0.013-bit spread (§4.4). In Finnish the two are level. At baseline
$H_{\mathrm{syll}}$ exceeds $H_{\mathrm{align}}$ by 0.010 bits, no more
than the 0.010-bit spread of the five settings. With sonority plateaux
allowed, the difference is 0.000 bits. In Italian the direction depends
on the transcription. With eSpeak-ng's symbols, $H_{\mathrm{syll}}$
exceeds $H_{\mathrm{align}}$ by 0.079 bits, beyond Italian's 0.025-bit
spread. With /ɪ/ and /i/ merged (§3.4), the excess is 0.018 bits, within
that spread. With the merge and sonority plateaux allowed,
$H_{\mathrm{syll}}$ lies 0.007 bits below $H_{\mathrm{align}}$. Of the
other seven languages, $H_{\mathrm{syll}}$ exceeds $H_{\mathrm{align}}$
in Norwegian and Swedish (by 0.186 and 0.161 bits) and falls below it in
Portuguese, French and Hungarian, each beyond its own spread and keeping
its sign under the Miller--Madow and jackknife corrections
(\texttt{run\_fall\_bias\_check.py}). Dutch's plug-in fall (0.020 bits)
does not survive the jackknife, which reads the two grains as level
(+0.004 bits). Danish is level (+0.004 bits, inside its 0.011-bit
spread). We read every direction off the plug-in value, and the
Miller--Madow and jackknife corrections move the three near-level
languages to small rises beyond their spreads: Danish by +0.031 and
+0.049 bits, Finnish by +0.0114 and +0.0125 (spread 0.0100) and Italian
with /ɪ/ and /i/ merged by +0.0263 and +0.0309 (its merged spread,
0.0253). English's sign depends on how its lexicon is built (§3.3). Over
the full dictionary, $H_{\mathrm{syll}}$ exceeds $H_{\mathrm{align}}$ by
0.167 bits. Over a uniform 20,000-word sample, the excess is 0.080 bits.
Over a frequency-matched 20,000-word list, $H_{\mathrm{syll}}$ is 0.015
bits below $H_{\mathrm{align}}$, so the plug-in sign differs between the
two 20,000-word lists. As §4.4 shows, this frequency-matched fall is
smaller than its leading-order bias correction. That sign also depends
on the aligner's treatment of /ju/: when ⟨u⟩ may spell /ju/ as one unit
(§2) the frequency-matched comparison is a 0.035-bit rise. \textbf{The
sign of that comparison carries little information on its own, and the
reason is mathematical.} A syllable contains several alignment units~---
2.23 to 2.61 of them across these five languages~--- so entropy measured
per syllable and entropy measured per alignment unit have different
denominators. If a block of $k$ units carried $k$ residual uncertainties
of $h$ bits each, independent of one another once the units' spellings
are known, the block's entropy would be exactly $kh$: larger per
syllable than per unit, with nothing having become less systematic. A
rise from the alignment-unit to the syllable scale can therefore occur
mechanically, from the unit's size alone. The comparison with $kh$ can
be made directly, by scoring each syllable's units one at a time against
the alignment-unit table: in every language the measured syllable
entropy falls below that summed score (English 1.10 against 2.40 bits
per syllable, the 5.84-bit per-word alignment-unit score below divided
by 2.43 syllables per word). So, in sample, a per-unit rise need not
reflect a loss of systematicity: the in-sample syllable table assigns
lower surprisal than the alignment-unit table gives the same units
scored one at a time, and that gain mixes the context a syllable takes
in with memorisation of sparsely observed syllables. §4.4 shows that
these in-sample figures are most sensitive to sparse observations at the
syllable grain.

The quantity comparable across the two grains is the \textbf{additive
conditional surprisal per word}, called the per-word score below. Let
the lexicon hold $W$ entries, let word $w$ contain $K_w$ units at grain
$g$, and let $\bar{K}_g$ be the mean units per word. Since the plug-in
entropy $\hat{H}_g$ (the $H_g$ of §2, computed from observed
proportions) averages the unit-level surprisal $-\log_2\hat{p}(y\mid x)$
over all $\sum_w K_w$ tokens, multiplying by $\bar{K}_g$ cancels the
token count and leaves

\[
\bar{K}_g\,\hat{H}_g=\frac{1}{W}\sum_w\sum_j -\log_2\hat{p}_g(y_{wj}\mid x_{wj}).
\]

In words: the product is the average, over words, of the summed
surprisals of that word's units, each scored against the pooled
conditional distribution for its grain.

\textbf{Additive conditional surprisal per word is an oracle score, not
a code length.} Each word's segmentation into units comes from the EM
alignment of that word's own pronunciation, so the number is scored with
the answer already in hand. A decoder holding only the spelling would
not know the segmentation, and nothing here charges it for finding one.
Two segmentations of one spelling condition on different events, so
their probability masses need not sum to one across that spelling's
possible pronunciations. In the smallest case, a spelling \texttt{ab}
read as a single unit and the same spelling read as \texttt{a} then
\texttt{b} can each receive probability 1 from one trained model, which
puts a total mass of 2 on the readings of a single spelling, summed over
its two segmentations, and violates Kraft's inequality (for a decodable
code, the probabilities $2^{-\ell}$ implied by its codeword lengths
$\ell$ sum to at most 1); each segmentation alone is normalised
(Appendix~\ref{appendix-a.-the-decomposition-of-the-per-word-fall}).
These numbers are therefore not bits anyone could transmit. They are
comparable across the two grains because the same kind of oracle, each
grain's own segmentation, is granted at each, which is all the
comparison below asks of them; the term $I$ below is the information the
spelling segmentation carries about the pronunciation, which the
syllable spelling discards. \textbf{The quantity is also not the
whole-word conditional entropy}, and the independence across units is a
property of the arithmetic rather than a claim about the language.

Under this unit-by-unit scoring the average per-word surprisal falls
from 5.84 to 2.68 bits in English and from 0.20 to 0.11 bits in Finnish
when the representation changes from alignment units to syllables. It
falls in all five reported languages, and in all twelve the released
table covers.

\textbf{\boldmath Only the spelling-boundary term $I$ can reverse the
per-word fall, and here it is small.} For any lexicon, per syllable
block, the difference between the two scores decomposes exactly
(Appendix A; Figure~\ref{fig:decomp}); the five lexicons determine only
the sizes of its terms. Write $X$ for the spellings of the block's
alignment units, $Y$ for their pronunciations, $G$ for the syllable
spelling and $P$ for its whole pronunciation (from here on, $G$ and $P$
denote the syllable's spelling and pronunciation, so $H(P\mid G)$ is the
syllable entropy $H_{\mathrm{syll}}$), and $L_f$ (f for fine) for the
block's expected summed alignment-unit surprisal. Then

\[
L_f-H(P\mid G)=D+H(Y\mid P,X)-I(P;X\mid G).
\]

In words: the alignment-unit score exceeds the syllable entropy by three
terms. $D$, the within-syllable term, is the context and dependence
inside the syllable that scoring its units one at a time ignores,
together with in-sample memorisation; $H(Y\mid P,X)$, the
pronunciation-boundary term, is the information lost about where the
pronunciation's internal boundaries fall; $I$, which is subtracted, is
the information the spelling segmentation carries about the
pronunciation, which the syllable spelling discards. $D$ and
$H(Y\mid P,X)$ are never negative, and $H(Y\mid P,X)$ is zero in all
five lexicons. Multiplying by syllables per word gives the lexicon's
mean per-word fall, which in these five lexicons therefore occurs
\textbf{\boldmath if and only if $D$ exceeds $I$} (in general, if and
only if $D+H(Y\mid P,X)$ exceeds $I$; Appendix A).

Only $I$ can make coarsening raise the score, and it is positive only
where one syllable spelling arises from several segmentations: the
syllable spelling is built by concatenating the units' spellings, which
discards the boundaries between them, and \textbf{1,473 of English's
27,162 syllable spellings, 5.42\%, arise from more than one distinct
fine segmentation} (German 2.30\%, Italian 1.67\%, Spanish 0.37\%,
Finnish 0.24\%). So the fall is not forced in general, and Appendix A
gives a two-word lexicon in which coarsening raises the score. In these
lexicons, though, $I$ is small. Each column of
Table~\ref{tab:decomp-terms} is a mean over the lexicon's words, the
per-block term times syllables per word (2.43 in English).

\begin{table}[htbp]
\centering
\caption{The lexicon's mean per-word fall in additive conditional
surprisal from alignment unit to syllable, with its within-syllable term
$D$ and spelling-boundary term $I$, each per
word.}\label{tab:decomp-terms}
\begin{tabular}{@{}llll@{}}
\toprule
language & per-word fall & $D$ per word & $I$ per word \\
\midrule
English & 3.158 & 3.358 & 0.200 \\
German & 2.623 & 2.735 & 0.112 \\
Italian & 1.415 & 1.426 & 0.011 \\
Spanish & 1.525 & 1.527 & 0.002 \\
Finnish & 0.083 & 0.085 & 0.002 \\
\bottomrule
\end{tabular}
\end{table}

Averaged over a lexicon's words, $I$ is largest in English, 0.20 bits
against English's 3.16-bit fall, then 0.112 bits in German and at most
0.011 bits in Italian, Spanish and Finnish; these are lexicon means, not
a bound on any single word. \textbf{The fall's positive sign therefore
rests on the identity plus two measured terms:} by the identity, a
lexicon's mean per-word rise is at most its mean per-word $I$, which is
at most 0.2004 bits here (English; 0.200 in
Table~\ref{tab:decomp-terms}), and the fall itself is the measured $D$
minus $I$. $D$ is in-sample: a syllable seen once is scored at zero
surprisal, so $D$ inherits the sparse-observation sensitivity §4.4
describes. Values come from
\texttt{pipeline/run\_grain\_decomposition.py}, which computes each term
by its own route.

\textbf{So per-unit entropy and additive per-word surprisal need not
move together as grain changes: the per-word score falls in all twelve
languages, while per-unit entropy falls in German and Spanish, is level
in Finnish, rises in Italian only with eSpeak-ng's symbols, and in
English depends on the lexicon.} Anyone comparing per-unit entropy
across grain sizes, as a difficulty index or otherwise, is comparing
quantities with different units; none of these entropies is offered as a
measure of what a reader computes.

\textbf{Against the nearest published direction.} Schmalz et al.~(2026,
pp.~10--11, Table 5) find per-unit conditional entropy falling in all
eight of their languages when the unit grows from the letter to the
bigram, over monosyllables. Of the seven languages both studies cover,
our per-unit change from alignment unit to syllable also falls in
French, German and Dutch (Dutch not under the jackknife). Under our
syllabification convention (§3.4), it is level in Danish by the plug-in
but rises under both corrections. It depends on the word list in English
and rises in Norwegian and Swedish. The designs differ at both ends.
Their smaller unit is the letter aligned one-to-one with phonemes, which
they note misaligns multi-letter spellings (p.~11), while ours is an
aligned unit that may span several letters; their larger unit is the
bigram of a monosyllable, ours the syllable of a polysyllabic word. We
have not tested which difference produces the disagreement in Norwegian,
Swedish and Danish.

\textbf{The per-word fall is not evidence that larger units are more
systematic.} Enlarging a unit does two things at once: it changes the
size of the thing being predicted, and it moves context from outside the
conditioning event to inside it. A two-word toy shows the second. Let a
language hold only the two word types \texttt{AB} and \texttt{AC},
pronounced /p r/ and /q s/. At the alignment-unit grain \texttt{A} is
ambiguous while \texttt{B} and \texttt{C} are not, so
$H_{\mathrm{align}}$ is 0.5 bits per unit over two units a word, which
is 1 bit per word; at the syllable grain each block is deterministic and
the figure is 0. Every letter maps exactly as before, so the whole fall
is context included in the larger unit: $D$ is 1 bit per word and $I$ is
0. In a real lexicon $D$ mixes that context with in-sample memorisation,
and the decomposition does not separate the two. We make no claim about
reading difficulty here either: the per-word figure is an in-sample
oracle score, and §4.4 records why it is sensitive to sparse
observations at the syllable scale.

\hypertarget{observed-unit-support-is-confounded-with-lexicon-size}{%
\subsubsection{3.3 Observed unit support is confounded with lexicon
size}\label{observed-unit-support-is-confounded-with-lexicon-size}}

$N$ is far from saturating at the syllable grain (coverage below), and
our English lexicon holds about 5.7 times as many word types as the
others, so raw counts of distinct syllable units are not comparable
without a common sample size. We match English at 20,000 words in two
ways~--- a uniform random sample of the dictionary, which matches size
only, and the 20,000 most frequent dictionary words in the OpenSubtitles
list, which matches the way the other four lexicons were built
(Table~\ref{tab:size}).

\begin{table}[htbp]
\centering
\caption{Syllable-spelling types and syllable entropy with each lexicon
at full size and matched at 20,000 words; only English's two rows
change.}\label{tab:size}
\begin{tabular}{@{}llll@{}}
\toprule
language & $N_{\mathrm{syll}}$ (full) & $N_{\mathrm{syll}}$
(matched) & $H_{\mathrm{syll}}$ (matched) \\
\midrule
English, uniform sample & 27,162 & 9,508 & 1.005 \\
English, frequency-matched & 27,162 & 9,084 & 0.879 \\
German & 6,252 & 6,252 & 0.510 \\
Italian & 2,690 & 2,690 & 0.510 \\
Spanish & 2,965 & 2,965 & 0.249 \\
Finnish & 2,889 & 2,889 & 0.039 \\
\bottomrule
\end{tabular}
\end{table}

\begin{figure}[!htbp]
\centering
\refstepcounter{figure}\label{fig:3}
\includegraphics[width=0.92\textwidth,height=0.7\textheight]{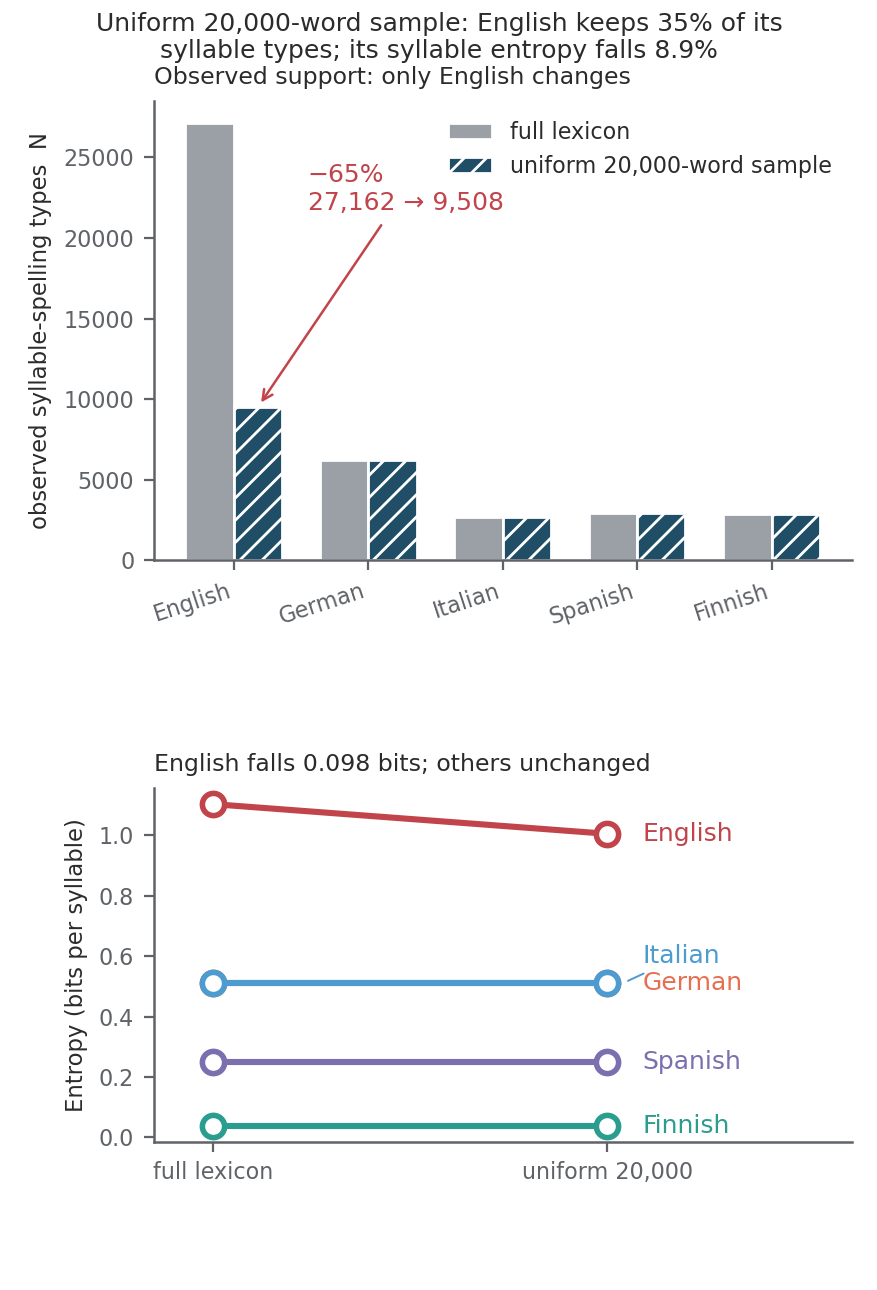}
\par\medskip
\begin{minipage}{\linewidth}
\textbf{Figure 4}~--- Top: observed syllable-spelling support
$N_{\mathrm{syll}}$ at the full lexicon and at the uniform 20,000-word
sample, per language. Only English moves, from 27,162 to 9,508 types,
because it is the only lexicon above the cap. Bottom: the paired
syllable-scale entropies, where English falls 1.103 → 1.005 bits while
the other four are unchanged by construction. The contrast between the
panels is the point: the uniform sample keeps only 35\% of English's
observed syllable-spelling types (9,508 of 27,162) yet lowers its
syllable entropy by only 8.9\%, so the two quantities are not equally
sample-dependent. German and Italian (0.5104 bits per syllable each,
equal to four decimals) carry separated labels (all markers are open);
their difference does not establish an ordering.
\end{minipage}
\end{figure}

\textbf{English is the only real contrast here}, because it is the only
language whose lexicon exceeds the 20,000-word cap; the other four rows
are unchanged by construction. The uniform row varies corpus size alone,
within one language (Figure~\ref{fig:3}). The frequency-matched row
varies composition too. It moves English's syllable entropy a further
0.126 bits below the uniform row, and 0.224 bits below the full
dictionary (0.138 with eSpeak-ng English, §2).

\textbf{\boldmath Entropy is less sample-sensitive than $N$, and it is
not invariant.} Matching English by uniform sample moves $H$ by 0.011
bits (1.1\%) at the alignment-unit grain and 0.098 bits (8.9\%) at the
syllable grain~--- the syllable scale is \textbf{9.3 times more
corpus-sensitive in bits}, and 7.9 times more as a share of its own
entropy (ratios from the unrounded shifts, 0.0106 and 0.0984 bits). That
matters for §3.2: English's own gap
$H_{\mathrm{syll}}-H_{\mathrm{align}}$ \emph{halves} under the uniform
match, from 0.167 to 0.080 bits. So in this uniform draw the gap is
about half its full-dictionary size. Over 1,000 further uniform
20,000-word draws (seeds 1 to 1,000) the matched gap averages 0.073
bits, with a standard deviation of 0.0056 and the central 95\% of draws
between 0.063 and 0.084, and it is positive in every draw. The halving
therefore does not depend on seed 31, whose 0.080 is at the 87th
percentile of the 1,000. These draws hold the dictionary and its
alignments fixed and vary only which words are sampled, so the spread
measures word sampling within this dictionary, not uncertainty about
English spelling as a whole
(\texttt{run\_entropy\_sampling\_spread.py}). With eSpeak-ng English
(§2) the syllable shift is 0.083 bits and the gap goes from 0.169 to
0.087, again about half. \textbf{Under a composition match, per-unit
entropy does not establish English's sign.} In the uniform sample
entropy still rises from the alignment-unit to the syllable grain in
English, as it does in Italian with eSpeak-ng's symbols, and still falls
in German and Spanish. In the frequency-matched list, without the /ju/
permission, $H_{\mathrm{align}}$ is 0.894 bits and $H_{\mathrm{syll}}$
0.879 bits. That 0.894 is the list's own value; it rounds to the same
figure as the full dictionary's under the permission (§3.4), a different
quantity. The list therefore shows a 0.015-bit per-unit fall. The
uniform sample shows a 0.080-bit rise. If ⟨u⟩ may spell /ju/ as one unit
(§2), the frequency-matched list shows a 0.035-bit rise. Under the same
permission, the uniform sample shows a 0.121-bit rise. So per-unit
entropy does not establish English's sign, the status §3.1 gives the
German--Italian order. Cut further by frequency, English's syllable
entropy falls below its alignment-unit entropy, by 0.091 bits at 10,000
words and 0.189 at 5,000. German, Italian and Spanish, whose alignments
are stored in frequency order, keep their signs at 5,000 and 10,000
words and over their full lists, and Finnish's change stays within 0.007
bits of zero at both cuts (0.010 over its full list, §3.2). The tests
were one uniform draw of 20,000 English words and frequency cuts: at
5,000 and 10,000 words in all five languages, at 20,000 for English, and
at the full lists of about 19,900 words for the other four. English is
the only one of the five whose direction changed under any cut we
tested. Any claim comparing unit counts across languages must compare
them at a common lexicon size. The matched column supplies one only by
reducing English, since the other four rows are their full lists.

\textbf{\boldmath $N$ counts observed units, not a cognitive inventory.}
$N$ counts the distinct spelling units that actually occur in the
lexicon we analyse. Matching lexicon size removes the gross asymmetry
between a 114,000-word English list and a 20,000-word one. That removal
is \textbf{\boldmath necessary before $N$ could be read as a cognitive
inventory in this design, and nowhere near sufficient}. The uniform
English row of the matched column uses a sample of $m$ word types drawn
uniformly \textbf{without replacement} from the lexicon's $M$ types. Its
expected support is therefore a rarefaction over a finite lexicon~---
ecology's sample-based rarefaction, with word types as the sampling
units and spelling units as the species (Colwell, Mao \& Chang, 2004,
Eq.~5)~--- and not an occupancy count under repeated independent draws
(draws with replacement, uniform over word types). Below, $C(a,b)$
counts the ways to choose $b$ items from $a$, and is zero when $b$
exceeds $a$. Let $c_u$ be the number of word types that contain unit $u$
at least once. After drawing $m$ of the $M$ types the expected number of
distinct units observed is

\[
\mathbb{E}[N_m]=\sum_u\Bigl(1-\frac{C(M-c_u,m)}{C(M,m)}\Bigr).
\]

In words: $C(M-c_u,m)/C(M,m)$ is the chance that a sample of $m$ types
misses every word carrying $u$, so one minus it is the chance $u$ has
been seen, and adding those chances across units gives the expected
count. The same expression gives the expected number of new units in the
next word:

\[
\mathbb{E}[N_{m+1}]-\mathbb{E}[N_m]=\sum_u\frac{c_u}{M-m}\cdot\frac{C(M-c_u,m)}{C(M,m)}.
\]

In words again: each unit contributes its chance of being still unseen,
times the chance the next draw reaches one of the $c_u$ words that carry
it.

\textbf{\boldmath One condition is necessary and sufficient for both the
strict rise and the shortfall (the expected number of observed units the
sample misses) at a given $m<M$: that some unit occurs in at most $M-m$
of the $M$ word types.} It is sufficient because any such unit has a
positive chance of being missed, which makes the shortfall positive and
the next word's expected yield positive with it; it is necessary because
if every unit occurs in more than $M-m$ types, every unit is certain to
appear in the sample, the shortfall is exactly zero and the curve has
stopped rising. Concavity needs no further hypothesis: term by term, for
every $m$ from 0 to $M-2$, each unit's next first difference equals its
current one times $(M-c_u-m)/(M-m-1)$, a factor at most 1 because
$c_u\geq 1$ (a term whose factor would be negative is already zero). So
the expected count never falls, rises strictly at every $m$ that meets
the condition above, and the rate of new discoveries never rises. Unlike
an infinite-population occupancy model, the curve reaches its ceiling:
at $m=M$ the sum is exactly the observed support, since a sample of the
whole lexicon sees every unit in it. Two languages matched on $m$ can
also differ in expected $N$ because their $c_u$ distributions differ, so
the matched column compares \emph{coverage at 20,000 words} rather than
inventory size.

\textbf{A realised plateau is not saturation.} A strict shortfall at
every sample size is not automatic: in a lexicon of the two words
\texttt{ab} and \texttt{ba}, one draw already shows every unit. It holds
for all twelve lexicons analysed here, because each contains at least
one alignment unit that occurs in a single word type (one in Finnish,
119 in English), and such a unit satisfies the condition above for every
$m<M$. So at every sample size $m<M$, at both grains (each lexicon also
holds syllable-spelling types that occur in a single word type, from
1,102 in Finnish to 16,094 in English), the expectation stays strictly
below what the whole lexicon shows; the grains differ in how far below,
which the next paragraph measures, and an observed flattening of $N$ is
a point on a discovery curve and not evidence that the unit system has
been fully seen. Establishing saturation would need a coverage or
unseen-support analysis, which we do not report. Every $N$ in this paper
should be read as observed support at the stated corpus size.

\begin{figure}[!htbp]
\centering
\refstepcounter{figure}\label{fig:4}
\includegraphics[width=0.92\textwidth,height=0.7\textheight]{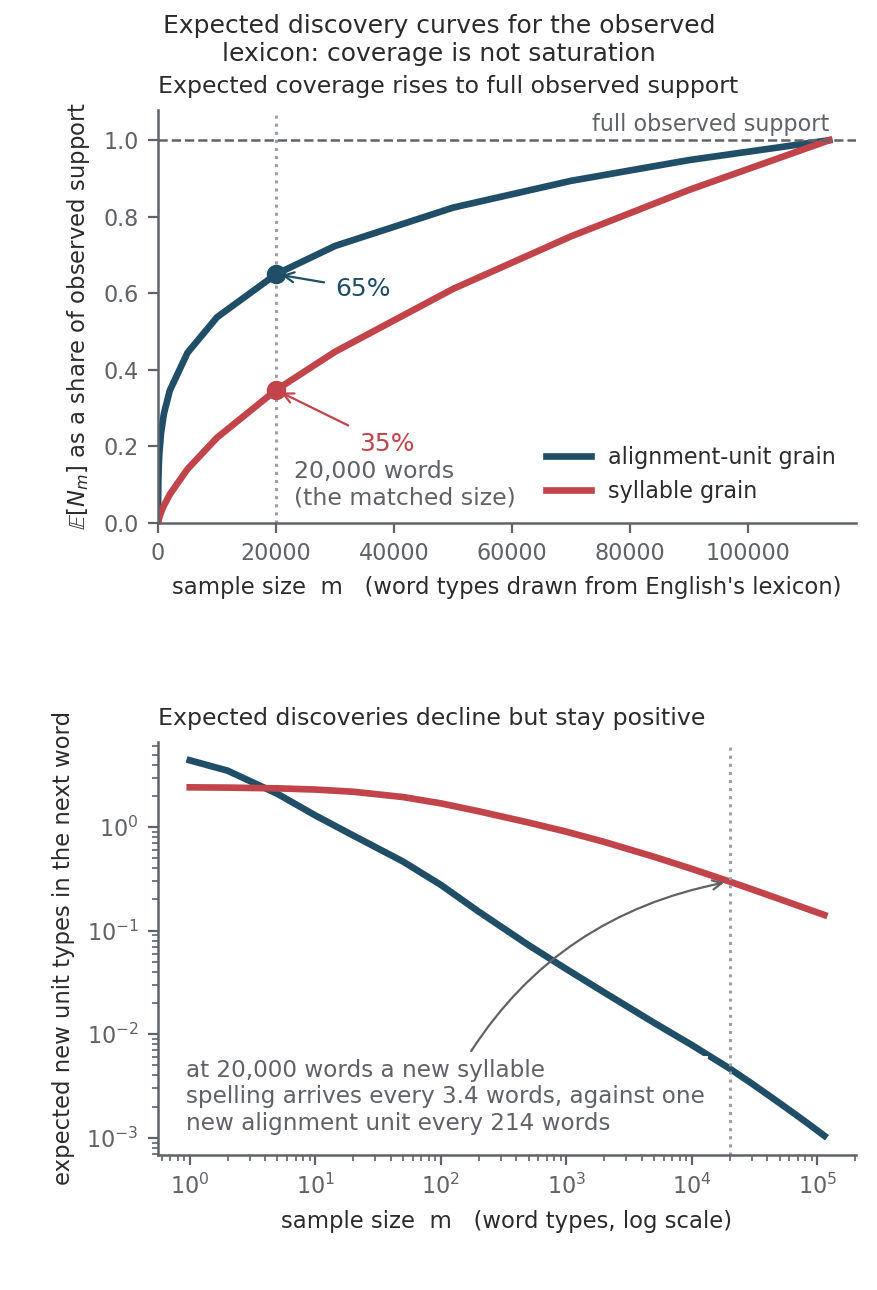}
\par\medskip
\begin{minipage}{\linewidth}
\textbf{Figure 5}~--- An intuition figure for the rarefaction
$\mathbb{E}[N_m]=\sum_u\bigl(1-C(M-c_u,m)/C(M,m)\bigr)$ with $c_u$
\textbf{measured from the English lexicon}, not assumed: $c_u$ is the
number of word types containing unit $u$ at least once, and $C(a,b)$
counts the ways to choose $b$ items from $a$, zero when $b$ exceeds $a$.
Top: expected observed support against sample size $m$, as a fraction of
the support the full lexicon exposes, at both grains. Both curves rise,
both are concave, and both meet the dashed line only at the last word,
where a sample of the whole lexicon has by definition seen every unit in
it. The dotted line marks the 20,000-word matched size, where the
alignment-unit grain has reached 65\% of its observed support and the
syllable grain 35\%. Bottom: the expected number of new unit types the
next word contributes, which stays strictly positive at every size short
of the full lexicon~--- at 20,000 words a new syllable spelling still
arrives every 3.4 words. \textbf{\boldmath Because $c_u$ is counted on
the same lexicon, each curve is a rarefaction of the support already
seen and cannot reach past it}~--- which only strengthens the point:
even restricted to units already observed, a 20,000-word sample is
expected to reach only 35\% of the syllable grain.
\end{minipage}
\end{figure}

\textbf{The matched column is far from full coverage.} Measuring $c_u$
directly from the English lexicon~--- the number of word types that
contain unit $u$ at least once~--- and evaluating the rarefaction above
gives the share of the observed support a sample of $m$ words is
expected to reach. At $m$ = 20,000, that is \textbf{65\% of the
alignment-unit spellings but only 35\% of the syllable spellings}
(Figure~\ref{fig:4}). At the same point a new syllable spelling arrives
every 3.4 words, against one new alignment unit every 214. The syllable
grain is not near the end of its discovery curve at the size every other
lexicon here is capped to.

\textbf{At the syllable grain the rarefaction is about 0.3 points from
the observed sample.} The uniform English sample contains 9,508
(Table~\ref{tab:size}) of the 27,162 observed syllable-spelling types,
which is 35.0\%. The rarefaction predicts 34.7\%, about 0.3 percentage
points below it; the independent-draws occupancy expectation predicts
32.8\%, which is 2.2 points low. A further 1,000 uniform 20,000-word
samples (seeds 1000 to 1999) give the scale: their syllable counts
spread with a standard deviation of 55 types (0.20 points), so the
sample sits 1.6 standard deviations above the rarefaction and the
occupancy expectation sits 9.1 standard deviations below the
rarefaction, and at this grain the sample tells the two models apart.
\textbf{At the alignment-unit grain the agreement is weaker.} The same
sample holds 331 of the 532 observed alignment-unit spellings (62.2\%)
against the 65.0\% the rarefaction predicts, 1.9 standard deviations low
(exact standard deviation 7.7 units, 1.45 points); 37 of the 1,000
further samples held 331 units or fewer. The occupancy expectation,
63.5\%, lies only 1.0 standard deviation below the rarefaction there, so
one draw cannot separate the two models at that grain. The sample is
within two standard deviations of the rarefaction at both grains, high
at one and low at the other. This comparison uses one realised draw
against an expectation, not a distributional test, and it confirms only
that the rarefaction model is consistent with the procedure that drew
the uniform English sample. \texttt{run\_rarefaction\_check.py} prints
every number in this paragraph, with a control that must fail and does:
the first 20,000 words in the dictionary's alphabetical order fall 25
standard deviations short at the syllable grain.

\hypertarget{robustness}{%
\subsubsection{3.4 Robustness}\label{robustness}}

\textbf{Syllabification.} We ran a baseline (maximum onset length of 3
consonants, /s/-cluster exception on, strict sonority) and four variants
of it: maximum onset 2, maximum onset 4, /s/-exception off, and sonority
plateaux allowed. The largest entropy shift any variant produces in
these five languages is \textbf{0.025 bits}, 2.3\% of the 1.065-bit
spread between English and Finnish at the syllable scale. English stays
highest and Finnish lowest under every variant. \textbf{The full
ordering does not survive every variant}: under maximum onset 2 and
under sonority plateaux allowed, German moves above Italian. That is
§3.1's four-decimal tie resolving one way, and it is the one place the
spread across syllabifier settings exceeds a between-language
difference.

\textbf{One coupling these five syllabifier settings cannot test, and
its size.} The syllabifier works on whole alignment units, so a boundary
falling inside a unit is never proposed and therefore never refused;
varying the onset rules cannot probe it. We bound it instead of testing
it. Every intervocalic consonant chunk spanning more than one phoneme
(under §2's constraint only ⟨x⟩, and ⟨z⟩ in German) is a place where the
ban \emph{could} refuse a boundary, and counting those gives \textbf{an
upper bound of 0.59\% of English's 215,919 intervocalic consonant
chunks} (German 0.25\%, Spanish 0.49\%, Italian 0.06\%, Finnish 0.03\%).
\textbf{That is a ceiling on how many boundaries the ban could block,
not a count of boundaries it did block}, and even the ceiling is under
one chunk in a hundred.

\textbf{Lone intervocalic consonants.} Giving every lone consonant
between two vowels to the preceding syllable, an extreme reversal of the
rule all five settings share (§2), shifts syllable entropy by up to 0.12
bits (English, down), almost five times the 0.025-bit spread above,
enough to put German above Italian and to bring Italian's rise (§3.2) to
level (−0.019 bits, inside its 0.025-bit spread). Counting only
doubled-letter consonants in both syllables, as ambisyllabic (⟨Hum⟩ +
⟨mer⟩, with Spanish ⟨ll⟩ and ⟨rr⟩ kept whole because each spells a sound
its single letter does not, and Spanish never divides it), moves the
five by at most 0.035 bits (Italian, up), leaves English highest and
Finnish lowest, puts Italian above German (0.546 against 0.510), outside
the 0.025-bit band, and reverses no sign of §3.2 among the five. Outside
the five, the same variant turns Dutch's jackknife reading from level
(+0.004 bits) to a fall (−0.067) and Danish's plug-in difference from
level to a fall (−0.020), while the jackknife still reads a Danish rise
(+0.025) and Miller--Madow a level difference (+0.007).

\textbf{Pronunciation source, English.} In the source-matched check of
§2 the forced /ju/ misalignment costs more than in the dictionary: it
raises English $H_{\mathrm{align}}$ by 0.066 to 0.072 bits (the low end
over all the words the corrected alignment covers, the high end over the
original run's words), against 0.042 to 0.046 in the dictionary. For the
next four values we apply the /ju/ permission to both sources and count
each source over all the words it aligns. English $H_{\mathrm{align}}$
is 0.919 bits with eSpeak-ng. With the dictionary it is 0.894 bits.
English $H_{\mathrm{syll}}$ is 1.088 bits with eSpeak-ng. With the
dictionary it is 1.101 bits. The tables, which use the dictionary
without the permission, print 0.936 and 1.103. End to end, switching the
source to eSpeak-ng, with its own retrained alignment and aligned word
set, raises English's alignment-unit entropy by 0.025 bits and lowers
its syllable entropy by 0.013. English stays highest and Finnish lowest
of the five at both grains, under each of the five syllabifier settings
above. Of the twelve, English is third at the alignment-unit grain
(second if Danish's stød marker is deleted) and highest at the syllable
grain over its full, unmatched lexicon (§3.1). The Caravolas calibration
of §3.1 is unchanged (19 of 21 pairs, rho = −0.80). Every per-unit sign
of §3.2 and §3.3 is the one the dictionary gives under the same
permission. So are the per-word fall and its decomposition: $I$, the
spelling-boundary term of §3.2, averages 0.129 bits per word, against
0.200 for the dictionary without the permission
(Table~\ref{tab:decomp-terms}). Without the permission, eSpeak-ng
English and Portuguese differ by 0.0004 bits at the alignment-unit grain
(0.9855 against 0.9851), too little to order them, so the Caravolas
count then turns on that gap (21 of 21 if it is read as an order, which
this paper does not do), and eSpeak-ng English lies 0.0057 bits above
Danish (0.9855 against 0.9798), where the dictionary puts it 0.044 bits
below. All three are /ju/ effects and go once the permission is applied.
One result depends on the source: the first-unit position check (§3.5),
with or without the permission, which is given both ways there. The
composition-matched English--Danish syllable comparison of §3.1 does
not: both sources put Danish above English (0.984 against 0.879 bits
with the dictionary, and against 0.951 with eSpeak-ng under the
permission). A few effect sizes move, and are also given both ways.

\textbf{Transcription artefacts.} eSpeak-ng encodes allophonic detail.
It transcribes the Finnish grapheme ⟨i⟩ as /ɪ/ in 72\% of its
occurrences in the word list and /i/ in 28\%. Suomi, Toivanen and
Ylitalo (2008, p.~20) give Finnish eight vowel phonemes with a single
/i/ among them. In these transcriptions the split follows syllable
position, not vowel length: ⟨i⟩ and ⟨ii⟩ are /i/ in every first-syllable
token and /ɪ/ in 90\% of later ones. So the two symbols are one phoneme,
and the transcription records allophonic detail as conditional
uncertainty the language does not have. Lyytinen et al.~(2015, p.~331)
put Finnish grapheme--phoneme consistency at 100\% in both directions,
which agrees. \textbf{We therefore merge the pair throughout}: every
table and figure, and every Finnish entropy quoted outside this
paragraph, uses the merged Finnish transcription, and the unmerged one
ships in the deposit. Merging lowers Finnish $H_{\mathrm{align}}$ by
\textbf{0.0655 bits, from 0.0947 to 0.0292}~--- 69\% of its unmerged
value~--- and $H_{\mathrm{syll}}$ from 0.141 to 0.039 bits, so the
per-unit rise of 0.046 bits under eSpeak-ng's symbols becomes 0.010
bits, within the spread of the five syllabifier settings (§3.2). It
raises the English--Finnish absolute difference from 0.8413 to 0.9068
bits. Expressed as a ratio the same change reads 9.9× to 32.1×, and we
report the absolute difference first and the ratio second, because
entropy is bounded below by zero and a ratio against a denominator this
close to it is unstable: a further 0.0001-bit shift in Finnish would
move the ratio by 0.11 while moving the absolute difference by 0.0001.
\textbf{Finnish is quoted to four decimals here for that reason}~--- at
three, 0.029 gives 32.3× and 0.0292 gives 32.1×, and the ratio inherits
a rounding the difference does not.

The same output contrast appears in Italian and German. In Italian,
which has no /ɪ/ phoneme, the pattern \textbf{is consistent with} a
shared transcription convention; in German, once length is stripped, /ɪ/
against /i/ is also the phonemic lax--tense contrast (⟨mit⟩ against
⟨Miete⟩), so there the pair can carry real orthographic uncertainty as
well as any convention. It does not establish the convention;
language-specific phonological review would. We therefore leave German
unmerged, report Italian both ways (the tables keep eSpeak-ng's symbols;
merging lowers Italian's $H_{\mathrm{align}}$ from 0.432 to 0.350 bits
and its $H_{\mathrm{syll}}$ from 0.510 to 0.368), and release the
candidate list for that review. Finnish alone is merged in the tables,
because only there do a phonology source and the position split above
both show the two symbols to be one phoneme.

\textbf{Choices we did not vary, and which way each could move the
orders.} For the eleven languages other than English we used no second
pronunciation source: eSpeak-ng transcribed every one of those lists,
and we looked none of their words up in a pronouncing dictionary.
Hand-built dictionaries exist for some of them (Schmalz et al., 2026,
p.~4, used CELEX for German and Dutch, the NST lexicon for the
Scandinavian languages and Lexique 2 for French), so for those languages
the check is possible and was not done. For all eleven, the four
reported in full (German, Italian, Spanish and Finnish) among them, the
transcription layer is unmeasured. A dictionary could move an eSpeak-ng
entropy either way: its lexical exceptions, which deterministic rules
suppress, would raise it, and dropping allophonic splits such as
Finnish's would lower it (§4.4). For English, the one language measured
both ways, the switch left every order intact once the /ju/ permission
was applied to both sources; without it, eSpeak-ng English and
Portuguese are 0.0004 bits apart and not ordered, and eSpeak-ng English
lies 0.0057 bits above Danish, which the dictionary puts above English
(Pronunciation source, English, above). The aligner was fitted once,
from one starting point, and its two chunk constraints (§2) were never
relaxed. A refit from another starting point could settle on other
segmentations and move any language's entropy in either direction. The
one alignment change we did measure, the /ju/ permission, moved
English's dictionary $H_{\mathrm{align}}$ by 0.042 to 0.046 bits,
changing no order between languages but reversing English's
frequency-matched per-unit sign (§3.3), and its eSpeak-ng
$H_{\mathrm{align}}$ by 0.066 to 0.072 bits, taking eSpeak-ng English
from first of the twelve, level with Portuguese, to third, below
Portuguese and Danish. Relaxing the constraints would let a unit absorb
neighbouring letters. For a fixed prediction target, absorbing them
moves context inside the unit, and added context cannot raise
uncertainty (§3.2's two-word toy), but a larger unit is also a larger
target, with more to predict, and §3.2 finds per-unit entropy rising
from alignment unit to syllable in Norwegian and Swedish; we have
measured neither the net direction nor its size in any language. Either
choice can reverse an order only by shifting two languages apart by more
than the gap between them. Per alignment unit, English lies 0.33 bits
above German and Finnish 0.29 bits below Spanish, the gaps behind the
five-language extremes; German lies 0.17 bits above Italian; English
lies 0.044 bits below Danish and 0.049 bits below Portuguese, the gaps
behind its third place among the twelve; and Finnish lies 0.017 bits
below Hungarian. So the extremes are the orders least exposed to either
choice, and English's third place and Finnish's lowest place among the
twelve are the most exposed.

\hypertarget{external-validity-checks}{%
\subsubsection{3.5 External validity
checks}\label{external-validity-checks}}

The statement about outcome prediction in §1 (contribution 1) and §4.4
rests on the Seymour, position and Ziegler analyses, with the Caravolas
calibration beside them, four in all, reported in
Table~\ref{tab:external} with their methods rather than asserted in
passing. In every row, ``our measure'' is the alignment-unit entropy
$H_{\mathrm{align}}$. All four use published cross-language data
(reading outcomes in two, reading-predictor weights in one, depth bands
in one) from which neither $H_{\mathrm{align}}$ nor onset entropy is
computed, and all four are exploratory rather than prespecified.
Spearman rank correlations throughout; code in \texttt{pipeline/} as
named.

\begin{table}[!htb]
\small
\centering
\caption{The four external validity checks of §3.5: for each, the
script, the published outcome measure, the languages compared and the
statistics.}\label{tab:external}
\begin{tabular}{@{}>{\raggedright\arraybackslash}p{0.19\linewidth}>{\raggedright\arraybackslash}p{0.21\linewidth}>{\raggedright\arraybackslash}p{0.14\linewidth}>{\raggedright\arraybackslash}p{0.37\linewidth}@{}}
\toprule
check (script) & outcome measure & languages & statistic \\
\midrule
Published depth bands (\texttt{run\_caravolas.py}) & Caravolas (2022,
p.~330) indicative bands & 8 that she places & 19 of 21 cross-band pairs
ordered correctly, p = 0.014 by 20,000-permutation label shuffle; rho =
−0.80, p = 0.010; with Danish's stød marker deleted (§3.1), 20 of 21 and
rho = −0.87; the same 19 of 21 and rho = −0.80 with eSpeak-ng English
(§2) \\
\addlinespace
Our measure ($H_{\mathrm{align}}$) against onset entropy
(\texttt{run\_seymour.py}) & Seymour et al.~(2003) Grade 1 familiar
content-word reading accuracy (their Table 5, the P1 rows, end of the
first school year) & 6~--- the whole intersection of Borgwaldt's seven,
Seymour's groups and ours & onset entropy rho = −0.66 (n = 6; exact
one-sided p = 0.088, two-sided 0.175), all-position rho = −0.31 (n = 6;
one-sided p = 0.282, two-sided 0.564, the same with eSpeak-ng English,
§2), exact over all 720 orderings \\
\addlinespace
Position, holding corpus, aligner and units fixed
(\texttt{run\_position.py}) & The same Seymour et al.~(2003)
outcome & 11 & $H_{\mathrm{align}}$ over non-initial alignment units rho
= −0.75 (exact two-sided p = 0.010), over all units (the all-position
measure) rho = −0.66 (n = 11; two-sided p = 0.031), over each word's
first alignment unit only rho = −0.64 (two-sided p = 0.040), exact over
all 39,916,800 orderings; with eSpeak-ng English (§2) −0.75 (p = 0.010),
−0.66 (p = 0.031) and −0.61 (p = 0.052) \\
\addlinespace
Second outcome set (\texttt{run\_ziegler.py}) & Ziegler et al.~(2010)
published phonological-awareness slope weights (their Table 2,
p.~555) & 5, of which only Finnish is among the five this paper
reports & onset entropy agrees better on three of their four outcomes
and ties on decoding speed (reading speed, reading accuracy, decoding
speed, decoding accuracy; a positive rho is agreement, higher entropy
with a larger phonological-awareness weight): rho +0.80, +0.90, +0.67
and +0.97 against our +0.70, +0.50, +0.67 and +0.67 \\
\bottomrule
\end{tabular}
\end{table}

\textbf{The hypothesis the Seymour and Ziegler checks were built to test
is not supported, and the position check leaves it open.} Landerl et
al.~(2013) decline entropy as a continuous predictor because onset
entropy misses irregularity outside the word onset, and an all-position
measure might have answered that. These comparisons do not show that it
does. Across the six languages both measures cover, Borgwaldt's onset
entropy has the stronger negative rank correlation with Seymour et
al.'s~(2003) Grade 1 familiar content-word accuracy (onset entropy rho =
−0.66, n = 6) than our all-position measure (rho = −0.31, n = 6). These
are correlations between language means, and Seymour et al.~(2003,
abstract and p.~167) attribute familiar-word differences mainly to
orthographic depth, and the slower start in English (as in Danish) to
syllabic complexity and depth combined, so neither measure's correlation
isolates depth. \textbf{The head-to-head cannot be widened past those
six}, and it cannot include Finnish: Borgwaldt et al.~measured Dutch,
English, French, German, Hungarian, Italian and Portuguese, so Finnish
is absent from their set, and neither Finnish value that circulates in
this literature is a Borgwaldt measurement: Landerl et al.~(2013, p.~688
and Table S1) supply an upper bound on the number of word-initial
letter-to-phoneme mappings (at most 21), not an entropy, and Ziegler et
al.~(2010, Table 1) an onset entropy of 0.0 computed from their own
Finnish corpus (p.~553). A seventh language would rest on one of those
cells, so we quote the six. The Ziegler check below does admit Ziegler
et al.'s Finnish cell, because it uses the onset-entropy column those
authors analysed beside their own outcomes (their Table 1: their corpus
value for Finnish, Borgwaldt et al.'s for the other four); the Seymour
head-to-head, by contrast, is against Borgwaldt's measure itself.

\textbf{The same direction appears in a second published outcome set},
and the second set is weaker corroboration than its four outcomes
suggest. Against Ziegler et al.'s (2010) published
phonological-awareness slope weights, onset entropy (Borgwaldt et al.'s
values for four languages, and Ziegler et al.'s own corpus value for
Finnish, their p.~553) agrees better than our measure on three of their
four outcomes and ties on the fourth, decoding speed (rho = +0.67 for
both). But those four outcomes come from the same five languages and the
same 1,265 children, and in Ziegler et al.'s Table 2 three of the four
phonological-awareness weight vectors are near-collinear: reading speed,
reading accuracy and decoding speed correlate at Pearson r = 0.92 to
0.99 across the five languages, while decoding accuracy departs from
them (r = 0.40 to 0.67). \textbf{These are not four independent tests.}
At five languages the exact one-sided permutation test has a floor of
1/120 = 0.008 without ties (2/120 = 0.017 for the two decoding outcomes,
which carry tied weights), but the binding constraint is the coarseness
above that floor: for every one of these four outcomes the p-value grid
holds at most two rungs below 0.05, and its third rung is 0.067, so one
swapped pair of languages carries a result across the conventional
threshold. Two of the four onset-entropy comparisons do fall below
it~--- p = 0.042 for reading accuracy and p = 0.017 for decoding
accuracy, both printed in the released comparison table~--- so a small p
is attainable at this sample size, and it is not what this check is
short of. And of those five languages only Finnish is among the five the
present paper reports. We record it as corroboration and never as
replication.

Two further points bear on the hypothesis: one fails to support it, and
one leaves it open. First, the stronger behavioural correlation comes
from Borgwaldt's first-letter-to-first-phoneme onset entropy despite its
restricted positional scope. Its authors had already extended it to two-
and three-letter contexts across the five languages of the 2004 study,
finding that ``the pattern of relative ambiguity of the five languages
as compared to one another remains about the same'' (Borgwaldt et al.,
2004, p.~178). That extension widened the word-initial context but kept
the first phoneme as the target, so it does not test whether measuring
non-initial positions adds anything. Our same-corpus position comparison
addresses that separate question. Second, on the eleven languages
Seymour et al.~cover (our twelve less Hungarian) the point estimates do
not favour the word's initial position. Holding corpus, aligner and
units fixed and varying only which part of the word is measured, the
correlation with Grade 1 accuracy is strongest for the
\textbf{non-initial} units (rho = −0.75, n = 11), against −0.66 for the
all-position measure (n = 11; equal by coincidence to onset entropy's
six-language rho) and −0.64 for initial units alone. On these point
estimates the rest of the word predicts the outcome at least as well as
the initial units do, though we report no test of the difference between
the correlations. The differences in rho (non-initial minus initial
−0.11, non-initial minus all-position −0.09) are descriptive, not tested
advantages. Deleting Danish's stød marker (§3.1) moves each of these
three correlations by at most 0.01. Their order is unchanged, and so it
is with English transcribed by eSpeak-ng (§2), which leaves the other
two correlations at their dictionary values. The first-unit p is 0.040
with the dictionary and 0.052 with eSpeak-ng, with or without the /ju/
permission, so it falls below 0.05 under one source only; the other two
stay below 0.05 under both.

\textbf{Our measure's association differs with the language set, by an
amount six languages cannot resolve.} The all-position measure
correlates with Grade 1 accuracy at rho = −0.31 over the six languages
shared with Borgwaldt and rho = −0.66 over all eleven Seymour covers~---
the same measure and outcome, and a difference that a six-language
estimate is too imprecise to establish. Covering more of the word did
not predict reading outcomes better than Borgwaldt's onset entropy in
these checks, and this paper does not claim it does. At n = 6 neither
measure's correlation rejects zero association: onset entropy's exact
one-sided p is 0.088, and ours is 0.282. The head-to-head therefore
compares two estimates neither of which rejects zero association, and we
read it as a failure to support the hypothesis, not as evidence against
it. At n = 11 our measure's correlation does reject zero association,
with an exact two-sided p of 0.031 (the same with eSpeak-ng English, §2)
over all 39,916,800 orderings of the eleven languages, uncorrected for
the three position measures tested against the same outcome. A Holm
adjustment over those three tests gives 0.030 for the non-initial
measure and 0.061 for the all-position and first-unit measures, under
either English source, so only the non-initial association clears 0.05
after adjustment, and clearing it does not make that measure the better
predictor. Onset entropy cannot be set against it there, because
Borgwaldt et al.~measured only six of those eleven languages.

\hypertarget{discussion}{%
\subsection{4. Discussion}\label{discussion}}

The three findings and §3.5's validity checks each matter for reading
research (§4.1). §4.4 sets out their limitations, §4.5 reports an
exploratory check, and §5 concludes.

\hypertarget{implications-for-reading-research}{%
\subsubsection{4.1 Implications for reading
research}\label{implications-for-reading-research}}

\textbf{Name the grain for any contrast in the middle of the range.}
English is highest and Finnish lowest of the five at both grains, under
either English pronunciation source and with the length mark restored
(§3.1, §3.4), as in first-year accuracy: the Scottish English-speaking
group read 33\% of familiar content words correctly, Finnish children
98\% (Seymour et al., 2003, Table 5, p.~153). Some mid-range contrasts
change with the grain. German lies 0.17 bits above Italian per alignment
unit and level with it per syllable (§3.1), or above it at both grains
if Italian's two i symbols are merged; Portuguese lies above English per
alignment unit and below it per syllable, and so does Danish while stød
is marked, against English's full dictionary. Matched on size and
composition, English lies below Danish per syllable too (0.879 against
0.984 bits, §3.1). A study using a mid-range depth difference as a
predictor must name the grain. German, Italian and Spanish children read
94\% to 97\% of their familiar content words correctly, and per
alignment unit these three languages reverse the order depth predicts
(higher entropy, lower accuracy), a three-language pattern we do not
test: German, the most accurate of the three (97\%), has the highest
entropy.

\textbf{Test the effect of a larger unit in each language.} Ziegler and
Goswami (2005, p.~11) write that smaller grain sizes tend to be more
inconsistent than larger ones, ``at least in English''. Per-unit entropy
falls from alignment unit to syllable in Spanish and, less securely
(§4.4), German, is level in Finnish, rises in Italian only with
eSpeak-ng's symbols, and in English depends on the word list (§3.2,
§3.3). Schmalz et al.~(2026) find it falling from letter to bigram in
all eight of their languages; of the seven languages both studies cover,
ours falls in three, French, German and Dutch (Dutch not under the
jackknife, §3.2). A cross-grain comparison should state unit, word list
and denominator: per unit, a larger unit can raise entropy through its
size alone; per word, the score falls in all twelve languages but is an
oracle score (§3.2). Our syllable is not Ziegler and Goswami's rime, so
their claim stays untested.

\textbf{Match lexicons on size and composition.} Cutting English to
20,000 words lowers its syllable entropy by 0.098 bits on average over
1,000 uniform draws (size alone) and by 0.224 bits under a frequency
cut, which changes composition too (§3.3). The larger shift exceeds the
full-lexicon English--Danish gap (1.103 against 0.984 bits, §3.1) and is
about the size of the Spanish--Finnish gap (0.210 bits,
Table~\ref{tab:depth}). Syllable-scale comparisons should match
word-type count, list construction, unit and pronunciation source, and
state the matched size. Corpora differing several-fold in size, such as
Schmalz et al.'s (2026) 1,155 to 8,027 monosyllables, need the same
matching, though we did not test their measures.

\textbf{Measuring every position did not predict better than the onset.}
Landerl et al.~(2013, p.~688) set entropy aside because Borgwaldt's
measure is ``based only on word onsets''. Over the six languages both
measures cover, onset entropy correlated with first-year accuracy at rho
= −0.66 and our all-position measure at −0.31; at n = 6 neither rejects
zero association, and no test compared the two (§3.5). Over the eleven
languages Seymour et al.~cover, ours is −0.66 (two-sided p = 0.031),
equal to onset entropy's six-language value only by coincidence. A
continuous depth measure needs testing in more than six languages; §4.5
reports an exploratory check over eleven.

\hypertarget{cue-reading-cipher-reading-and-neighbouring-fields}{%
\subsubsection{4.2 Cue reading, cipher reading and neighbouring
fields}\label{cue-reading-cipher-reading-and-neighbouring-fields}}

Philip Gough distinguished cue reading, in which a beginner recognises a
word by a salient visual cue in or around it, from cipher reading, which
replaces it once the child decodes through letter--sound relations
(Gough \& Hillinger, 1980; Gough, Juel, \& Griffith, 1992, as cited in
Ehri, 2005, p.~138). Children read their first 40 or so words by cues,
and the strategy breaks down because visual cues run out (Gough \&
Hillinger, 1980, as cited in Ehri, 2005, p.~142). In those terms the
conditional entropy describes the cipher itself, not cipher reading: how
variable its correspondences are across a lexicon. It does not measure a
learner's uncertainty while inferring them, nor the effort of learning
them. Nor is an orthography a cipher in the cryptographic sense, which
its keyholder can invert exactly: a reader who knows English spelling
well still cannot tell from ⟨ea⟩ alone whether it is read as in
\emph{bead} or \emph{head}. We use the analogy to mark what the measure
covers, not as a formal result.

\textbf{Cryptography.} Shannon (1949) separated a cryptanalyst's
uncertainty about the key, the key equivocation, from uncertainty about
the message, the message equivocation. Our aligner works as a
cryptanalyst handed known plaintext: it receives spellings paired with
their pronunciations and infers the correspondences between them. In
those terms the fitted correspondence table plays the key, and the
entropy we report is the uncertainty about the message that remains once
the key is known. For a true cipher that quantity is zero. For English
it is not, and it says nothing about how hard the table is to learn.

\textbf{Coding theory.} Coding theory first fixes what each decoder is
given. Slepian and Wolf (1973) derived the coding rates, in bits per
source character, needed for two correlated sources under sixteen
arrangements of what each encoder and decoder has access to. In those
terms our per-word score is a decoder handed the pronunciation's
syllable boundaries, and a reading model must instead say how boundaries
are recovered from print. We do not identify the score with a
Slepian--Wolf rate.

\textbf{Grapheme-to-phoneme conversion.} Speech technology predicts a
pronunciation from a spelling and scores the prediction on words
withheld from training. Joint-sequence models (Bisani \& Ney, 2008)
learn the alignment and the probabilities together from training words,
then predict the pronunciations of words withheld from training. Fitting
such a model with the training vocabulary, the alignment constraints and
the handling of unseen units all fixed in advance would test whether the
grain-dependent patterns here hold on unseen words. Our alignments were
learned on every word, so that test needs a new fit and is left to later
work (§4.5).

\textbf{Learning models.} Connectionist models make spelling--sound
consistency matter through learning. The networks of Plaut, McClelland,
Seidenberg and Patterson (1996) reproduced the interaction of word
frequency and spelling--sound consistency, in naming latency for their
recurrent network, through how often each mapping was met in training
and how spelling and sound were represented. Our estimates describe
distributions such a model could learn from, under stated choices of
vocabulary and unit, with every word type weighted equally. They do not
assume that readers represent the aligner's units or compute its
entropy. Weighting each word by its frequency, which the deposited
alignments do not carry, would show how much the weighting choice
matters.

\hypertarget{granularity-and-transparency}{%
\subsubsection{4.3 Granularity and
transparency}\label{granularity-and-transparency}}

Wydell and Butterworth's hypothesis has two dimensions, and these
measurements bear on one. Transparency, how close to one-to-one the
print-to-sound translation is at whatever level it operates (their
p.~280), is a property of the writing system, estimated here from word
lists at two scales. Their Table 2 (p.~280) places Italian, German,
English and Danish along it in that order, from transparent to opaque,
and the twelve-language alignment-unit entropies order the four the same
way (0.432, 0.604, 0.936 and 0.980 bits), Danish above English only
while its stød marker is kept (§3.1). At the syllable scale the order
does not hold: German and Italian tie (0.510 bits each; merging
Italian's two i symbols puts it lower, at 0.368), and Danish (0.984)
falls below English's full dictionary (1.103), though not below English
matched on size and composition (0.879) (§3.1). Granularity in their
sense, the size of the smallest orthographic unit representing sound (at
the coarse end, a whole character or word), is a property of the script.
It does not vary among alphabetic orthographies and is untested here.
The grain we vary is our own unit of analysis inside an alphabetic
script, not the script's granularity. The reader's grain lies outside
these measurements: whether a reader such as AS (Wydell \& Butterworth,
1999) has acquired the sub-syllabic mappings whose failure those authors
tie to phonological dyslexia (p.~280) is a fact about that reader.

\hypertarget{limitations}{%
\subsubsection{4.4 Limitations}\label{limitations}}

\textbf{Word lists and readers.} The measurements describe writing
systems as sampled by particular word lists. They do not describe
readers, and they do not explain dyslexia. We tested no readers. The
pronunciations are transcriptions, not any reader's phonology, and one
transcription convention moved Finnish's alignment-unit entropy by 69\%
of its uncorrected value until we corrected it (§3.4). The main results
cover only five European alphabetic languages. The lexicons are a
pronouncing dictionary and frequency lists from film subtitles, not the
vocabularies children meet while learning to read. \textbf{We do not
claim that broader positional coverage predicts reading outcomes better.
A direct validity check against Borgwaldt's onset entropy did not
support the hypothesis that broader coverage predicts better}, a
same-corpus position comparison leaves it open, and §3.5 reports both.

\textbf{Unit size and context.} Enlarging the unit necessarily includes
context, so reporting both grains in §3.2 does not separate the
contributions of unit size and context. §3.2 also interprets the
direction of per-unit entropy cautiously, for a reason that is
mathematical rather than empirical: the unit has changed size, so the
two numbers are not in the same units.

\textbf{Estimation, transcription and method.}

\begin{itemize}
\tightlist
\item
  \textbf{Finite-sample error, largest at the syllable grain.} With a
  fixed alignment, conditional entropy calculated from pooled
  alignment-unit tokens is the descriptive plug-in statistic of the
  observed lexicon. Sparse spelling units often contribute zero observed
  conditional entropy, which can produce substantial downward error: a
  spelling unit seen once receives empirical conditional entropy of
  exactly zero, which records that no second pronunciation was observed
  and not that none exists. For independent sampling of individual unit
  tokens, the plug-in conditional entropy is downward biased in
  expectation. Our sampling unit, however, is the complete word, and
  words contribute unequal numbers of unit tokens. Under that design the
  pooled estimator is a ratio statistic whose finite-sample bias does
  not have a guaranteed sign. Two equally likely word types, drawn
  independently, show it: one holds 14 tokens of a spelling unit, all
  read one way, and the other holds two tokens of it, read two ways.
  Pooled over tokens, the population value is 0.337 bits, yet the
  plug-in from one sampled word averages 0.5 bits and from two words
  0.419, both above it. The Miller--Madow and synthetic-lexicon checks
  below illustrate sensitivity to sparse observations but establish
  neither a bias direction nor a bound for the word-sampled estimator.
  \textbf{At the syllable grain, 38\% to 59\% of observed
  syllable-spelling types occur exactly once} (Finnish 38\%, English's
  full dictionary 59\%), against 1.5\% (Finnish) to 41\% (Italian) of
  alignment-unit spellings seen as a single token at the alignment-unit
  grain. The leading Miller--Madow correction (Miller, 1955, p.~97,
  eqs.~10 and 12), $(r-1)/(2t\ln 2)$ for a spelling unit seen $t$ times
  with $r$ readings and averaged over units by their share of unit
  tokens, is +0.0001 to +0.0008 bits at the alignment-unit grain and
  +0.002 to +0.024 bits at the syllable grain, in the five full
  lexicons. \textbf{That correction does not estimate the error, and the
  sparse case shows it can fall far short of it}: on synthetic lexicons
  built to have a true conditional entropy of exactly 1.000 bit, a dense
  one (50 units seen 2,000 times each) recovers 1.0000, while a sparse
  one (20,000 units seen twice each) reads 0.503 and the correction
  raises it only to 0.684~--- recovering 36\% of the 0.497-bit shortfall
  and leaving 0.316 bits uncorrected. We make no claim that the
  correction bounds the error from below. Miller's leading term counts
  the outcomes that have nonzero probability (his $k$, p.~95); applied,
  as is standard practice, with the observed number of readings in their
  place, it is a leading-order term in the \emph{observed} support
  count, and we supply no hypotheses under which it is a lower bound.
  The corrections quantify sensitivity under their assumptions and do
  not establish the sign of bias in the actual word sampling design. The
  volume's editor adds that Miller's formulas ``are not to be used when
  number of observations and number of categories are of similar sizes''
  (editor's note in Miller, 1955, p.~100), the case of a syllable
  spelling seen only once or twice. The full-dictionary sign of
  $H_{\mathrm{syll}}-H_{\mathrm{align}}$ in each of the five languages
  of §3.2 survives the correction, while English's frequency-matched
  per-unit fall does not (−0.015 becomes +0.019 bits), and the /ju/
  treatment of §2 alone also reverses it (to +0.035 bits). So the
  frequency-matched per-unit fall is smaller than the 0.034-bit shift
  its leading-order bias correction produces and establishes neither a
  fall nor a rise. German's and Spanish's falls also keep their sign
  under two stronger corrections: a delete-one-word jackknife, which
  reads 1.006 bits on the sparse synthetic lexicon above where
  Miller–Madow reads 0.684, gives −0.065 and −0.068 bits respectively,
  and extrapolating the plug-in to an unlimited lexicon gives −0.074 and
  −0.069 (\texttt{run\_fall\_bias\_check.py}). The check is weaker for
  German. With pronunciations planted on each language's real syllables
  so that the true syllable entropy lies 0.05 bits above
  $H_{\mathrm{align}}$, every estimator detects the planted rise in
  Spanish, but in German none reads it as a rise beyond German's
  0.0131-bit spread: the jackknife reads the two grains as level (+0.004
  bits) and the others read a fall. Under that planted model, adding
  each estimator's planted shortfall to its German estimate leaves a
  fall beyond that spread only for the jackknife (−0.020 bits), while
  Miller--Madow and extrapolation read level (−0.0129 and −0.0111 bits).
  Under the Miller–Madow correction, at the syllable grain, German lies
  above Italian (0.528 against 0.520), but inside Italian's 0.025-bit
  spread, so the correction no more orders the two than baseline does
  (§3.1). \textbf{No bracket on the population value is claimed.} For a
  fully prespecified probability model (an aligner trained only on the
  training words and a prespecified backoff that gives every unseen
  syllable some probability), the population cross-entropy, the expected
  surprisal under the same token-weighted joint distribution,
  upper-bounds the population conditional entropy. A realised held-out
  score only estimates that expectation, and neither its value nor, when
  words of unequal length are sampled, its own expectation is guaranteed
  to lie above the population entropy. Nor does the bound furnish a
  finite-sample bracket here, because the in-sample plug-in has no
  guaranteed bias sign under word sampling. Our in-sample figures, whose
  alignment is learned on the same words, are claimed as a bound in
  neither direction, and no held-out cross-entropy is computed here.
\item
  \textbf{The entropy is an in-sample corpus statistic.} The alignment
  is learned from the same words on which entropy is computed, so the
  figures describe the analysed lexicon. Multiple EM restarts and a
  cross-validated aligner would estimate the effect of fitting the
  alignment on the same words, and are not reported; §3.4 states which
  way a refit could move each order.
\item
  \textbf{Automatic phonemization may bias entropy in either direction,
  and the direction is language-specific.} Deterministic rules can
  suppress lexical exceptions, which lowers entropy; allophonic detail
  can introduce distinctions that are not phonemic for the target
  language, which raises it. §3.4 documents one instance of the second
  kind, which moved Finnish's alignment-unit entropy by 69\% of its
  unmerged value. Because neither direction nor magnitude is general,
  cross-language comparison needs phonemic normalization or
  source-matched sensitivity analysis.
\item
  \textbf{English's pronunciations come from a different source than the
  other four.} CMUdict against eSpeak-ng, as §2 states. For English it
  is measured (§2, §3.4): compared end to end, with the /ju/ permission
  applied to both sources and each alignment retrained on its own
  transcription, the switch leaves every ranking, the Caravolas
  calibration and every per-unit sign; switching the source moves one
  position p-value across 0.05 (0.040 with the dictionary, 0.052 with
  eSpeak-ng, with or without the permission). For the other four it is
  unmeasured, because we used no second pronunciation source for them
  (§3.4), so every cross-language comparison here compares transcription
  processes as well as orthographies: the depth measured is depth as
  these transcriptions represent it.
\item
  The two chunk constraints are linguistic assumptions made to defeat a
  likelihood length bias, not something the data forced, and we report
  no ablation for them; §3.4 states which way relaxing them could move
  the orders.
\item
  The syllabifier is a heuristic, and its baseline and four variants are
  a local perturbation analysis rather than a gold-standard validation.
  Syllable structure and phonotactic legality differ by language; a
  stratified comparison against a trusted syllabified resource per
  language would be better.
\item
  We report the reading direction only. Spelling-direction entropy is
  not covered here.
\item
  We have not separated orthographic from phonotactic unpredictability
  in general. §3.4 handles one instance; the general problem Schmalz et
  al.~raise requires per-language inspection and is unsolved here.
\item
  The aligner is a restricted reimplementation of established
  many-to-many EM methods (Jiampojamarn et al., 2007; the restrictions
  are §2's two chunk constraints), not a new algorithm. It is released
  for reproducibility, not as a contribution.
\item
  Apart from the Wilson intervals of the alignment check (§2), the
  spread of the 1,000 further uniform samples of syllable counts (§3.3),
  which scales one comparison, and the spread of English's entropy over
  1,000 uniform draws (§3.3), which holds the dictionary fixed, no
  inferential interval accompanies any entropy, count or correlation
  estimate above. Repeated subsampling of the other languages would give
  one. The rank correlations in §3.5 are computed over five to eleven
  languages, so their estimates are imprecise and sensitive to which
  languages are included. We report exact permutation p-values for them
  and no inferential intervals.
\item
  We located the prior work through the reference lists of the reviews
  and comparisons cited here and through targeted literature-database
  and web searches. It is not a systematic review.
\end{itemize}

\hypertarget{future-directions}{%
\subsubsection{4.5 Future directions}\label{future-directions}}

The limits above name their own next steps: a second pronunciation
source for the eleven eSpeak-ng languages (§3.4), a held-out
cross-entropy, repeated subsampling for languages other than English and
for lexicons matched on composition, a per-language comparison against a
trusted syllabified resource, and tests with readers. One check needs no
new data, and we report it below as exploratory.

\textbf{An exploratory check against first-year accuracy.} We ran it
after its inputs were in hand, so it is exploratory, not a prespecified
test, and its p-value carries no confirmatory weight. It covers the
eleven languages of §3.5's position check. English enters cut to its
20,000 most frequent words, transcribed by eSpeak-ng as the other ten
lists are, and aligned under the /ju/ permission of §2 (the letter u may
spell /j/ plus a vowel as one unit). Across the eleven, syllable-scale
entropy rank-correlates with Seymour et al.'s (2003, Table 5, p.~153)
Grade 1 (first-year) familiar content-word accuracy at rho = −0.74,
exact one-sided p = 0.006 over all 11! orderings: the languages whose
syllables leave more pronunciation uncertainty had lower first-year
accuracy. With the dictionary transcription, rho is −0.75 (p = 0.006)
for the uniform draw (seed 31) and −0.74 under the frequency cut, so the
English source and matching do not move the result. Two controls behave
as they must: reversing the accuracies gives rho = +0.74 (p = 0.99), and
placing English lowest drops the association to rho = −0.25 (p = 0.23),
so the association depends heavily on where the two extreme languages
fall. Its alignment-unit counterpart is already reported (rho = −0.66, n
= 11, §3.5), and Seymour et al.~group their languages by syllable
complexity, so the correlation may track syllable complexity rather than
depth. The two scales order the eleven languages too alike to separate
them. The check shows only that the syllable-scale value tracks this
outcome in these eleven languages. \texttt{run\_syllable\_accuracy.py}
recomputes it from the deposit.

\hypertarget{conclusion}{%
\subsection{5. Conclusion}\label{conclusion}}

We measured how predictably spelling maps to sound, as conditional
entropy over every aligned unit in every word of full polysyllabic
lexicons, in twelve alphabetic orthographies at two grain sizes, the
alignment unit and the syllable. English is highest and Finnish lowest
of the five reported in full at both grains. Per-unit entropy has no
common direction as units grow, and some mid-range contrasts change with
the grain: Portuguese lies above English per alignment unit and below it
per syllable, so a study using a mid-range depth difference as a
predictor must name the grain. Lexicon size changes the syllable-grain
estimate: a uniform 20,000-word sample of the English dictionary has
syllable entropy 0.098 bits lower, averaged over 1,000 draws whose
standard deviation is 0.006 bits, so syllable-scale comparisons should
match lexicons on size and composition. For reading research, the
measure gives cross-language studies a continuous depth value at a
stated grain, with every position in the word contributing rather than
the onset alone, though over six languages it tracked first-year
accuracy no better than onset entropy (§3.5), and §4.5 reports an
exploratory check in which its syllable-scale value tracked first-year
reading accuracy, an association that depends heavily on where English
and Danish fall.

AS's dissociation (§1) is a fact about one reader, and these estimates
describe the spelling-sound patterns a reader of each language meets.
English is highest of the five at both grains, which fits the
transparency half of Wydell and Butterworth's hypothesis, but a
word-list statistic cannot say which grain AS read by or why his
difficulty stayed in English. Because the estimates move with the grain,
the vocabulary and the transcription, those three choices belong in any
claim about orthographic depth. The exploratory association with
first-year accuracy establishes neither a cause of reading difficulty
nor an advantage for the syllable. Settling the granularity half needs
readers. One test would fit syllable-scale scores without the evaluation
words and ask whether they predict word-naming accuracy beyond
alignment-unit scores, word frequency and length. A failure would count
against a syllable advantage.

\begin{center}\rule{0.5\linewidth}{0.5pt}\end{center}

\hypertarget{references}{%
\subsection{References}\label{references}}

Bisani, M., \& Ney, H. (2003). Multigram-based grapheme-to-phoneme
conversion for LVCSR. In \emph{Proceedings of Eurospeech 2003}
(pp.~933--936). ISCA. https://doi.org/10.21437/Eurospeech.2003-326

Bisani, M., \& Ney, H. (2008). Joint-sequence models for
grapheme-to-phoneme conversion. \emph{Speech Communication, 50}(5),
434--451. https://doi.org/10.1016/j.specom.2008.01.002

Black, A. W., Lenzo, K., \& Pagel, V. (1998). Issues in building general
letter to sound rules. In \emph{Proceedings of the Third ESCA/COCOSDA
Workshop on Speech Synthesis} (pp.~77--80). ISCA.

Borgwaldt, S. R. (2003). \emph{From onset to entropy:
Spelling-pronunciation patterns in six languages} {[}Doctoral thesis,
Universiteit van Amsterdam{]}.
\url{https://pure.uva.nl/ws/files/3949148/49099_Thesis.pdf}

Borgwaldt, S. R., \& De Groot, A. M. B. (2002). Beyond the Rime:
Measuring the Consistency of Monosyllabic and Polysyllabic Words. In M.
Neef, A. Neijt, \& R. Sproat (Eds.), \emph{The Relation of Writing to
Spoken Language} (pp.~49--69). Niemeyer.
https://doi.org/10.1515/9783110918601.49

Borgwaldt, S. R., Hellwig, F. M., \& De Groot, A. M. B. (2004).
Word-initial entropy in five languages --- letter to sound and sound to
letter. \emph{Written Language and Literacy, 7}(2), 165--184.
https://doi.org/10.1075/wll.7.2.03bor

Borgwaldt, S. R., Hellwig, F. M., \& De Groot, A. M. B. (2005). Onset
entropy matters --- letter-to-phoneme mappings in seven languages.
\emph{Reading and Writing, 18}(3), 211--229.
https://doi.org/10.1007/s11145-005-3001-9

Borleffs, E., Maassen, B. A. M., Lyytinen, H., \& Zwarts, F. (2017).
Measuring orthographic transparency and morphological-syllabic
complexity in alphabetic orthographies: A narrative review.
\emph{Reading and Writing, 30}(8), 1617--1638.
https://doi.org/10.1007/s11145-017-9741-5 (read in the open-access full
text, Europe PMC PMC5574968, which carries no page numbers, so the
pinpoint names a section)

Caravolas, M. (2022). Reading and reading disorders in alphabetic
orthographies. In M. J. Snowling, C. Hulme, \& K. Nation (Eds.),
\emph{The science of reading: A handbook} (2nd ed., pp.~327--353).
Wiley-Blackwell. https://doi.org/10.1002/9781119705116.ch15

Carnegie Mellon University. (2008). \emph{The CMU Pronouncing
Dictionary} (Version 0.7a) {[}Data set{]}.
http://www.speech.cs.cmu.edu/cgi-bin/cmudict

Chee, Q. W., Chow, K. J., Yap, M. J., \& Goh, W. D. (2020). Consistency
norms for 37,677 english words. \emph{Behavior Research Methods, 52}(6),
2535--2555. https://doi.org/10.3758/s13428-020-01391-7

Colwell, R. K., Mao, C. X., \& Chang, J. (2004). Interpolating,
extrapolating, and comparing incidence-based species accumulation
curves. \emph{Ecology, 85}(10), 2717--2727.
https://doi.org/10.1890/03-0557

Damper, R. I., Marchand, Y., Marsters, J.-D. S., \& Bazin, A. I. (2005).
Aligning text and phonemes for speech technology applications using an
EM-like algorithm. \emph{International Journal of Speech Technology, 8},
147--160. https://doi.org/10.1007/s10772-005-2166-6

Dave, H. (2018). \emph{FrequencyWords} (2018 OpenSubtitles frequency
lists) {[}Data set{]}. GitHub.
https://github.com/hermitdave/FrequencyWords

Ehri, L. C. (2005). Development of sight word reading: Phases and
findings. In M. J. Snowling \& C. Hulme (Eds.), \emph{The science of
reading: A handbook} (pp.~135--154). Blackwell.
https://doi.org/10.1002/9780470757642.ch8

eSpeak NG contributors. (2024). \emph{eSpeak NG} (Version 1.52 for the
English transcriptions; earlier, unrecorded releases for the other
eleven lists, §2) {[}Computer software{]}.
https://github.com/espeak-ng/espeak-ng

Galescu, L., \& Allen, J. F. (2001). Bi-directional conversion between
graphemes and phonemes using a joint N-gram model. In \emph{Proceedings
of the 4th ISCA Tutorial and Research Workshop on Speech Synthesis}.
ISCA.

Jiampojamarn, S., Kondrak, G., \& Sherif, T. (2007). Applying
many-to-many alignments and hidden Markov models to letter-to-phoneme
conversion. In \emph{Proceedings of NAACL HLT 2007} (pp.~372--379).
Association for Computational Linguistics.
https://aclanthology.org/N07-1047/

Landerl, K., Ramus, F., Moll, K., Lyytinen, H., Leppänen, P. H. T.,
Lohvansuu, K., et al.~(2013). Predictors of developmental dyslexia in
European orthographies with varying complexity. \emph{Journal of Child
Psychology and Psychiatry, 54}(6), 686--694.
https://doi.org/10.1111/jcpp.12029

Lange, M., \& Content, A. (1999). The grapho-phonological system of
written French: Statistical analysis and empirical validation. In
\emph{Proceedings of the 37th Annual Meeting of the Association for
Computational Linguistics} (pp.~436--442). Association for Computational
Linguistics. https://aclanthology.org/P99-1056/

Lim, A., O'Brien, B., \& Onnis, L. (2024). Orthography--phonology
consistency in English: Theory- and data-driven measures and their
impact on auditory vs.~visual word recognition. \emph{Behavior Research
Methods, 56}(3), 1283--1313. https://doi.org/10.3758/s13428-023-02094-5

Lison, P., Tiedemann, J., \& Kouylekov, M. (2018). OpenSubtitles2018:
Statistical rescoring of sentence alignments in large, noisy parallel
corpora. In \emph{Proceedings of the Eleventh International Conference
on Language Resources and Evaluation (LREC 2018)}. European Language
Resources Association. https://aclanthology.org/L18-1275/

Lyytinen, H., Erskine, J., Hämäläinen, J., Torppa, M., \& Ronimus, M.
(2015). Dyslexia --- Early identification and prevention: Highlights
from the Jyväskylä Longitudinal Study of Dyslexia. \emph{Current
Developmental Disorders Reports, 2}(4), 330--338.
https://doi.org/10.1007/s40474-015-0067-1

Marjou, X. (2021). OTEANN: Estimating the transparency of orthographies
with an artificial neural network. In \emph{Proceedings of the Third
Workshop on Computational Typology and Multilingual NLP} (pp.~1--9).
Association for Computational Linguistics.
https://doi.org/10.18653/v1/2021.sigtyp-1.1

McCarthy, A. D., Lee, J. L., DeLucia, A., Bartley, T., Agarwal, M.,
Ashby, L. F. E., Del Signore, L., Gibson, C., Raff, R., \& Wu, W.
(2023). The SIGMORPHON 2022 shared task on cross-lingual and
low-resource grapheme-to-phoneme conversion. In \emph{Proceedings of the
20th SIGMORPHON Workshop on Computational Research in Phonetics,
Phonology, and Morphology} (pp.~230--238). Association for Computational
Linguistics. https://doi.org/10.18653/v1/2023.sigmorphon-1.27

Miller, G. A. (1955). Note on the bias of information estimates. In H.
Quastler (Ed.), \emph{Information theory in psychology: Problems and
methods} (pp.~95--100). Free Press. (read in the Internet Archive
lending copy, item informationtheor0000henr)

Plaut, D. C., McClelland, J. L., Seidenberg, M. S., \& Patterson, K.
(1996). Understanding normal and impaired word reading: Computational
principles in quasi-regular domains. \emph{Psychological Review,
103}(1), 56--115. https://doi.org/10.1037/0033-295X.103.1.56

Protopapas, A., \& Vlahou, E. L. (2009). A comparative quantitative
analysis of Greek orthographic transparency. \emph{Behavior Research
Methods, 41}(4), 991--1008. https://doi.org/10.3758/BRM.41.4.991

Schmalz, X., Marinus, E., Coltheart, M., \& Castles, A. (2015). Getting
to the bottom of orthographic depth. \emph{Psychonomic Bulletin \&
Review, 22}(6), 1614--1629. https://doi.org/10.3758/s13423-015-0835-2
(read in the author copy on OSF, osf.io/vxw8c)

Schmalz, X., Rueckl, J. G., \& Siegelman, N. (2026). How we should
measure orthographic depth: Or should we? \emph{Psychonomic Bulletin \&
Review, 33}, 45. https://doi.org/10.3758/s13423-025-02831-1

Seymour, P. H. K., Aro, M., \& Erskine, J. M. (2003). Foundation
literacy acquisition in European orthographies. \emph{British Journal of
Psychology, 94}(2), 143--174.
\url{https://doi.org/10.1348/000712603321661859}

Shannon, C. E. (1949). Communication theory of secrecy systems.
\emph{Bell System Technical Journal, 28}(4), 656--715.
https://doi.org/10.1002/j.1538-7305.1949.tb00928.x

Siegelman, N., Kearns, D. M., \& Rueckl, J. G. (2020). Using
information-theoretic measures to characterize the structure of the
writing system: The case of orthographic-phonological regularities in
English. \emph{Behavior Research Methods, 52}(3), 1292--1312.
https://doi.org/10.3758/s13428-019-01317-y

Slepian, D., \& Wolf, J. K. (1973). Noiseless coding of correlated
information sources. \emph{IEEE Transactions on Information Theory,
19}(4), 471--480. https://doi.org/10.1109/TIT.1973.1055037

Suomi, K., Toivanen, J., \& Ylitalo, R. (2008). \emph{Finnish sound
structure: Phonetics, phonology, phonotactics and prosody} (Studia
Humaniora Ouluensia 9). University of Oulu.

Torres, C. J., \& Futrell, R. (2025). \emph{Clarifying orthography:
Orthographic transparency as compressibility} (arXiv:2505.13657). arXiv.
https://doi.org/10.48550/arXiv.2505.13657

Van den Bosch, A., Content, A., Daelemans, W., \& De Gelder, B. (1994).
Measuring the complexity of writing systems. \emph{Journal of
Quantitative Linguistics, 1}(3), 178--188.
https://doi.org/10.1080/09296179408590015 (read in the authors'
preprint, whose page numbers the pinpoints give)

Wydell, T. N., \& Butterworth, B. (1999). A case study of an
English-Japanese bilingual with monolingual dyslexia. \emph{Cognition,
70}(3), 273--305. https://doi.org/10.1016/S0010-0277(99)00016-5

Ziegler, J. C., Bertrand, D., Tóth, D., Csépe, V., Reis, A., Faísca, L.,
Saine, N., Lyytinen, H., Vaessen, A., \& Blomert, L. (2010).
Orthographic depth and its impact on universal predictors of reading: A
cross-language investigation. \emph{Psychological Science, 21}(4),
551--559. https://doi.org/10.1177/0956797610363406

Ziegler, J. C., \& Goswami, U. (2005). Reading acquisition,
developmental dyslexia, and skilled reading across languages: A
psycholinguistic grain size theory. \emph{Psychological Bulletin,
131}(1), 3--29. https://doi.org/10.1037/0033-2909.131.1.3

\textbf{Note on Borgwaldt et al.~(2004).} Obtained and read at primary.
Both quotations above are from its page 178, read off the rendered page
rather than a text layer. The second author's initial is given as ``F.
M.'', from this article's own byline; the 2005 article's reference list
prints it as ``F.''

\textbf{Note on sources.} Every other cited work was also checked
against a held copy of the source for the use made of it here
(claim-support check, 2026-10-16; Colwell et al., Jiampojamarn et
al.~and Marjou on 2026-10-20; Chee et al.~and Miller on 2026-10-21,
Miller in full, pp.~95--100, in the Internet Archive lending copy; Ehri,
and the pages of Seymour et al., Ziegler and Goswami and Wydell and
Butterworth that §4 cites, on 2026-10-22; Lange and Content,
pp.~437--438, Borgwaldt et al.~(2004), pp.~167--169 and 182, and Schmalz
et al.~(2026), pp.~10--11, on 2026-10-22, and Borgwaldt and De Groot
(2002), pp.~49--56, on 2026-10-29, read as rendered pages), except Van
den Bosch et al.~(1994), read in the authors' preprint, and Schmalz et
al.~(2015), read in an author copy, both on 2026-10-20; Borleffs et
al. (2017), read in its open-access full text on 2026-10-20; and the
four entries for the data and software used (CMUdict, eSpeak NG,
FrequencyWords and OpenSubtitles2018), checked against their own
repositories or records.

\makeatletter\phantomsection\def\@currentlabel{A}\makeatother
\hypertarget{appendix-a.-the-decomposition-of-the-per-word-fall}{%
\subsection{Appendix A. The decomposition of the per-word
fall}\label{appendix-a.-the-decomposition-of-the-per-word-fall}}

\textbf{Setting.} Draw a syllable block uniformly from all syllable
tokens in a lexicon. Let $X=(X_1,\ldots,X_J)$ be its alignment units'
spellings and $Y=(Y_1,\ldots,Y_J)$ their aligned pronunciations; the
number of units $J$ varies from block to block and is fixed by $X$. Let
$G$ be the syllable spelling, the concatenation of the $X_i$, and $P$
its concatenated pronunciation. Let $q(y\mid x)$ be the pooled
alignment-unit table of §2, and let
$L_f=\mathbb{E}\bigl[\sum_i -\log_2 q(Y_i\mid X_i)\bigr]$, the block's
summed surprisal when its units are scored one at a time.

\textbf{Claim.} For the empirical block distribution of any lexicon,
exactly (the proof uses only that $J$ is fixed by $X$, $P$ is a function
of $Y$ and $G$ a function of $X$),

\begin{align*}
L_f-H(P\mid G)&=D+H(Y\mid P,X)-I(P;X\mid G),\\
D&=\mathbb{E}_X\,\mathrm{KL}\Bigl(p(Y\mid X)\,\Big\Vert\,\prod_i q(Y_i\mid X_i)\Bigr).
\end{align*}

In words: the alignment-unit score exceeds the syllable entropy by how
far the unit-by-unit predictor is from the lexicon's own (empirical)
distribution of a block's pronunciations, plus the
pronunciation-boundary information that concatenating the pronunciations
discards, minus the information the spelling segmentation carries about
the pronunciation, which the syllable spelling discards.

\textbf{Proof.} Because $J$ is fixed by $X$, the product
$\prod_i q(y_i\mid X_i)$ is a normalised distribution over pronunciation
sequences given $X$, so $L_f$ is a conditional cross-entropy, and a
cross-entropy is the entropy plus the divergence: $L_f=H(Y\mid X)+D$.
The whole pronunciation $P$ is a function of $Y$, so
$H(Y\mid X)=H(P\mid X)+H(Y\mid P,X)$. The syllable spelling $G$ is a
function of $X$, so $H(P\mid X)=H(P\mid X,G)=H(P\mid G)-I(P;X\mid G)$.
Substituting the third equation into the second, and the second into the
first, gives the claim. ∎

\textbf{From blocks to words.} Both scores in §3.2 use the same words,
every alignment unit lies in exactly one syllable, and the syllable
score is the plug-in $H(P\mid G)$ of the empirical block distribution.
So the alignment-unit score per word is $\bar{S}\cdot L_f$ and the
syllable score per word is $\bar{S}\cdot H(P\mid G)$, where $\bar{S}$ is
the mean number of syllables per word ($\bar{K}_g$ of §3.2 at the
syllable grain), and the per-word fall is $\bar{S}$ times the left side
of the claim.

\textbf{Sign.} $D$ and $H(Y\mid P,X)$ are never negative, and $I$ is
subtracted. The lexicon's mean per-word score falls under coarsening
\textbf{\boldmath if and only if $D+H(Y\mid P,X)>I$}. A syllable
spelling that determines its segmentation, so that $X$ is a function of
$G$, makes $I$ zero; that is sufficient for the score not to rise, and
it is not necessary. Without it the fall is not forced. Take two word
types, both spelled \texttt{AB} (homographs), one made of the units
\texttt{A} and \texttt{B}, read /x/ and /y/, and one made of the single
unit \texttt{AB}, read /z/. Every unit is deterministic, so the
alignment-unit score is 0, while the syllable spelling \texttt{AB} now
has two readings and the syllable entropy is 1 bit: $D$ is 0, $I$ is 1
bit, and coarsening raises the score by 1 bit per word. Homographs are
not needed: in the two words \texttt{ABC}, with units \texttt{A},
\texttt{B}, \texttt{C} read /k/, /a/, /e/ and syllables \texttt{AB} and
\texttt{C}, and \texttt{ABD}, with units \texttt{AB}, \texttt{D} read
/o/, /e/, every unit is again deterministic, the syllable spelling
\texttt{AB} has two readings, and coarsening again raises the score by 1
bit per word.

\textbf{Computation.} \texttt{pipeline/run\_grain\_decomposition.py}
computes $D$ as a divergence, $I$ as $H(X\mid G)-H(X\mid G,P)$ and
$H(Y\mid P,X)$ from triple counts, so the two sides of the claim are
reached by different routes; they agree to within 10⁻¹¹ bits in all five
languages. $H(Y\mid P,X)$ is zero in all five, a measured fact rather
than a structural one: in these alignments, a syllable's spelling units
and whole pronunciation always determine how the pronunciation divides
among the units. Two controls can fail. Dropping $I$ must leave a
residual equal to −$I$, and it does, nonzero in every language. A
syllable spelling rebuilt to keep the boundaries between units must give
$I$ exactly zero, and it does. The decomposition's terms, the
Miller--Madow figures of §4.4 and the exact tests of §3.5 were also
recomputed with separately written parsing, syllabification and entropy
code that imports nothing from \texttt{pipeline/}. That code is in the
deposit, in \texttt{independent/}. The two routes share the deposited
alignment files, and the phoneme-class tables, which are data.

\makeatletter\phantomsection\def\@currentlabel{B}\makeatother
\hypertarget{appendix-b.-deposit-contents-and-the-script-behind-each-number}{%
\subsection{Appendix B. Deposit contents and the script behind each
computed number}\label{appendix-b.-deposit-contents-and-the-script-behind-each-number}}

The deposit named in the availability statement holds the aligner, the
per-language alignments, the alignment-unit and syllable entropy tables
for all twelve languages, the lexicon-size control table
(Table~\ref{tab:size}), the syllabification sensitivity table, the
allophony candidate list (transcription symbol pairs that may record one
phoneme as two, §3.4) and the scripts named below. The pronunciations
behind the tables ship inside the alignments, but no deposited script
can regenerate the eSpeak-ng transcriptions without eSpeak-ng, and the
eleven non-English ones were made with a version that was not recorded
(§2), so a rerun may not reproduce them exactly.

\textbf{We do not release the frequency lists themselves} (the words
with their counts). The lists are derived from OpenSubtitles, so the
archive's \texttt{NOTICES.md} names the lists' source (FrequencyWords,
built from OpenSubtitles2018) and links it rather than copying it, and
\texttt{corpora.py} refetches each one with its counts. The word lists
do ship, inside the deposited alignments: each alignment file carries
every aligned word's spelling and pronunciation. The eleven non-English
files keep their lists' rank order, carry no counts, and are released
under the lists' own CC BY-SA 4.0 licence (\texttt{LICENSE-DATA.md}).
The English files (five alignments and \texttt{espeak\_en\_lexicon.tsv})
carry CMUdict's word forms, and CMUdict's own BSD-style licence is
reproduced in \texttt{NOTICES.md}. The deposit also carries the
twelve-language tables and the per-language chunk reports (each
language's 20 largest contributors to its entropy, with their commonest
pronunciations) the allophony candidates are read off.

\textbf{Recomputed from the deposited files alone.} With no network
access and no frequency list: the syllabification variants of §3.4
(\texttt{run\_syll\_sensitivity.py}); the Miller--Madow, singleton and
synthetic-lexicon figures of §4.4 (\texttt{run\_entropy\_bias.py}); the
decomposition and collision counts of §3.2 and the uniform
matched-control row of §3.3 (\texttt{run\_grain\_decomposition.py}); the
coverage figures of §3.3 and Figure~\ref{fig:1}'s per-unit contributions
and 77\% vowel share (\texttt{figures\_paper2.py}); the allophony
correction of §3.4, with Italian both ways and the per-unit change under
each syllabifier setting and, for Finnish and Italian, under the
Miller--Madow and jackknife corrections (\texttt{run\_allophony.py});
the released Finnish alignment, written from eSpeak-ng's unmerged
transcription, which ships as \texttt{transcription\_raw/align\_fi.txt}
(\texttt{run\_transcription\_merge.py}); the Danish stød values of §3.1
and §3.5 (\texttt{run\_allophony.py}, \texttt{run\_caravolas.py},
\texttt{run\_position.py}); the ceiling on blocked boundaries of §3.4
(\texttt{run\_boundary\_ceiling.py}); the twelve-language table's
alignment-unit and syllable rows, with how its word counts reconcile
(\texttt{run\_entropy\_table\_check.py}); the rarefaction comparison of
§3.3 at both grains (\texttt{run\_rarefaction\_check.py}); the alignment
check of §2, with its fixed sample and the reader's verdicts
(\texttt{run\_alignment\_audit.py}); the /ju/ comparison of §2, with the
re-aligned English file it reads (\texttt{run\_ju\_check.py}); the
source-matched English check of §2 and §3.4, with the eSpeak-ng
transcriptions and alignment it reads (\texttt{run\_espeak\_english.py};
its \texttt{-\/-rerun-pipeline} outputs ship in
\texttt{espeak\_rerun/}); the same check under the /ju/ permission of
§2, with the two re-aligned eSpeak-ng files it reads
(\texttt{run\_ju\_check\_espeak.py}; its \texttt{-\/-rerun} outputs ship
in \texttt{ju\_espeak\_rerun/}); the stronger bias corrections of §3.2
and §4.4, for all eleven eSpeak-ng languages
(\texttt{run\_fall\_bias\_check.py}); the single-occurrence counts of
§1, §3.3 and §4.4 (\texttt{run\_singleton\_counts.py}); the Finnish
position split of §3.4 (\texttt{run\_finnish\_position.py}); the
placeholder scan of §2 and its 201 German words
(\texttt{run\_placeholder\_scan.py}); English's entropy over 1,000
uniform 20,000-word draws, §3.3 and §4.4
(\texttt{run\_entropy\_sampling\_spread.py}); the exploratory check of
§4.5 against first-year accuracy, exact p-values and controls included
(\texttt{run\_syllable\_accuracy.py}); and the four checks of §3.5,
exact p-values included (\texttt{run\_caravolas.py},
\texttt{run\_seymour.py}, \texttt{run\_position.py},
\texttt{run\_ziegler.py}). All of them use only the Python standard
library except \texttt{figures\_paper2.py}, which needs
\texttt{matplotlib}. The second route of Appendix A ships beside them in
\texttt{independent/} (\texttt{p31\_g02\_recompute.py},
\texttt{p31\_independent\_check.py}); it needs \texttt{mpmath} and
imports nothing from \texttt{pipeline/}. The frequency-matched English
row of §3.3 and its §4.4 correction (−0.015 to +0.019) need the English
frequency list, which \texttt{run\_grain\_decomposition.py} reads once
\texttt{corpora.py} has refetched it. The length-mark comparison of §3.1
(0.558 against 0.530) needs eSpeak-ng, because the deposited alignments
have the length mark removed: \texttt{run\_lengthmark.py} re-phonemizes
each word and restores the mark onto the deposited segmentation wherever
the re-phonemized tokens match the deposited ones (all but 117 German
words), and the output it printed for this paper ships as
\texttt{lengthmark\_both\_ways.txt}. The ß comparison of §3.1 needs the
German frequency list, because the deposited alignments carry the
uppercased spellings: \texttt{run\_eszett\_check.py} reads the list once
\texttt{fetch\_sources.py} has fetched it, and its printed output ships
as \texttt{eszett\_check.txt}. Every other number the paper prints is
one of four kinds: a value quoted from a cited source, with its page; a
setting of the aligner or of a sample (§2); the CMUdict entry counts of
§2 (114,954 of 115,533 alphabetic entries), counted on the NLTK copy
that \texttt{corpora.py} reads; or a frequency-list figure (the share of
dictionary words absent from the list in §2, the frequency-matched row
and frequency cuts of §3.3, the §4.4 correction, and the 0.224 of §3.3
and §4.1), which \texttt{run\_grain\_decomposition.py} prints once
\texttt{corpora.py} has refetched the list.

\subsection*{Data and code availability}\addcontentsline{toc}{subsection}{Data and code availability}
The aligner, the per-language alignments, the alignment-unit and
syllable entropy tables for all twelve languages, the lexicon-size and
syllabification tables, the allophony candidate list and the scripts
named in Appendix B are released as
\texttt{p31\_paper2\_osf\_bundle.zip} in the OSF project at
\url{https://doi.org/10.17605/OSF.IO/A9T3C}. The frequency lists
themselves (the words with their counts) are not released: the deposit
names their source, and a deposited script refetches each one from it.
The words themselves ship, with their pronunciations, inside the
alignment files. Regenerating the eleven non-English pronunciations
needs eSpeak-ng, whose version for them was not recorded (§2), so a
rerun may not reproduce them exactly. Appendix B lists what the deposit
holds, which script recomputes each computed number, which numbers no
deposited script recomputes, and what the scripts need in order to run.

\subsection*{Use of AI tools}\addcontentsline{toc}{subsection}{Use of AI tools}
AI tools assisted under the author's direction. Claude (Anthropic) wrote
analysis and verification code to the author's specification and gave
drafting and editorial support. A separate Claude instance judged the
alignment-check sample of §2. ChatGPT (OpenAI) was used for adversarial
review and mathematical derivation, Perplexity for literature search and
citation verification, and Wolfram Mathematica for checking derivations.

\bigskip

\noindent\begin{minipage}{\linewidth}
\begin{verse}
\emph{Each grain of spelling holds a clue to sound,}\\
\emph{more context comes when separate grains are bound.}\\
\emph{Finnish speaks plainly, English veils its cues,}\\
\emph{the measure shifts with grain and words we choose.}
\end{verse}
\end{minipage}

\end{document}